\documentclass[a4paper]{article} % For LaTeX2e
\usepackage{amsmath}
\usepackage{amsfonts}
\usepackage{amssymb}
\usepackage[dvipsone]{graphicx}
\usepackage{subfig}
\usepackage{hyperref}
\newenvironment{acknowledgements}{%
  \begin{abstract}
}{%
  \end{abstract}
}
\def\pics{./}
\title{Auto-Associative Memories for Direct Signaling of Visual Angle During Object Approaches}

\author{
Matthias S. Keil\thanks{Also: \url{https://www.neurociencies.ub.edu}, Institute for Neurosciences
Edifici de Ponent, Campus Mundet, Universitat de Barcelona, Passeig Vall d'Hebron, 171. E-08035 Barcelona} \\
Department Cognition, Development and Psychology of Education\\
University of Barcelona\\
E-08035 Barcelona, Spain\\
\texttt{matskeil@ub.edu} \\
}

\begin{document}

\maketitle

\begin{abstract}
Being hit by a ball is usually not a pleasant experience. While a ball may not be fatal, other objects can be. To protect themselves, many organisms, from humans to insects, have developed neuronal mechanisms to signal approaching objects such as predators and obstacles. The study of these neuronal circuits is still ongoing, both experimentally and theoretically. Many computational proposals rely on temporal contrast integration, as it encodes how the visual angle of an approaching object changes with time. However, mechanisms based on contrast integration are severely limited when the observer is also moving, as it is difficult to distinguish the background-induced temporal contrast from that of an approaching object. Here, I present results of a new mechanism for signaling object approaches, based on modern content-addressable (auto-associative) memories. Auto-associative memories were first proposed by Hopfield in 1982, and are a class of simple neuronal networks which transform incomplete or noisy input patterns to complete and noise-free output patterns. The memory holds different sizes of a generic pattern template that is efficient for segregating an approaching object from irrelevant background motion.  Therefore, the model's output correlates directly with  angular size.  Generally, the new mechanism performs on a par with previously published models. The overall performance was systematically evaluated based on the network's responses to artificial and real-world video footage.  A gentle introduction to the key ideas of this paper is available on \href{https://youtu.be/LUNtzz4Mi8w}{Youtube}.
\end{abstract}
%
%
%--------------------------------------------------------------
\section{Introduction\label{SecIntro}}
%--------------------------------------------------------------
%
Detection of collision threats through visual information is vital for many organisms \cite{MatsJoan2012}. When an observer (e.g. a robot or an organism) does not move, tracking an approaching object is a straightforward computational exercise. However, when the observer is moving (and looking) straight ahead \cite{Gibson1950}, all objects in its field of view appear to collide. The additional movement due to the self-motion of the observer is referred to as \emph{background motion} or \emph{background movement}. With background motion, a computational challenge is to distinguish objects that will eventually collide with the observer from those that will pass by. In other words, any collision detection system should filter out background motion such that it responds to any colliding object in the same way as it would without background movement.\\
It is worthwhile to understand the neuronal circuitry of the locust's lobula giant movement detector (LGMD) neurons, as they reliably respond to approaching objects in depth, even in the presence of background motion \cite{Schlotterer1977,RindSimmons92a,RindSimmons92b}. Two types of LGMD neurons are distinguished by their responses to luminance contrast: LGMD1 responds to both lighter and darker objects than the background \cite{RindSimmons92a}, while LGMD2 responds only to darker objects \cite{SimmonsRind1997}. When probing the LGMD1's response to object approaches against a drifting grating, a reduction in response occurs \cite{RindSimmons92a}. For the LGMD2, the reduction is less pronounced for intermediate drifting frequencies\footnote{The (effective) drifting frequency of a grating increases both with its spatial frequency and temporal frequency.} \cite{SimmonsRind1997}.
The effect of background suppression on LGMD responses is further highlighted by showing locusts selected parts of the Star Wars movie \cite{RindSimmons92a} or dashcam videos showing car crashes and less harmful traffic footage \cite{Hartbauer17}.\\
Computational models for the LGMD usually start with calculating the difference between two consecutive (gray-level) video frames ($=$\emph{temporal contrast} or isotropic optical flow).  Temporal contrast extraction is a frame-rate based implementation of event-based signal processing in the sense that a signal is only generated if a movement occurs from one frame to the next \cite{TayaraniSchmuker21,GallegoEtAl2022}.

For an approaching object in the absence of background motion, the spatial sum of temporal contrast ($=$\emph{SOC}) correlates with the object's angular velocity (Figure \ref{FigApproachCurves}).  Angular velocity refers to the rate of change of the visual angle.   For driving LGMD responses,  (temporal) contrast edges were identified as a relevant feature \cite{RindSimmons92b}, when these edges increase in size and velocity in concert with an approaching object.  Activity related to temporal contrast provides excitatory input to the LGMD.\\
In parallel, inhibition to neighboring spatial positions in retinotopic space is generated (lateral inhibition for short).  If excitatory activity is eventually to build up in the LGMD neuron and trigger a response, then excitation must escape the inhibitory wavefront.  This occurs for approaching objects, because the closer the object gets, the bigger will be its image projected on the retina (angular size), and the faster its edges will grow (angular velocity)  \cite{RindBramwell96,RindSimmons98}.  Lateral inhibition therefore implements a predictive mechanism for non-approaching objects \cite{MatsEliAngel04}\\
Two further inhibitory mechanisms may act to avoid undesired responses and improve the suppression of background motion: \emph{(i)} large-field feedforward inhibition is activated upon a large increment of SOC from one time step to the next; this prevents corresponding activity from building up in the LGMD. For example, such sudden increases can occur in response to changes in the viewing direction and/or self-motion.
\emph{(ii)} The mean LGMD activity across the recent past can also be subtracted from the instantaneous LGMD activity.  Along with adequate thresholding of the LGMD response the baseline excitation due to background motion is removed.\\
If the inhibitory pathways have lowpass characteristics in time, then persistence of past activity in these channels decreases responsiveness and object approaches may be missed.
This problem can be solved by increasing the number of parallel inhibitory channels, such that the activity in each channel is kept low on the average. For example, temporal contrast can constitute parallel ON (positive values of temporal contrast) and OFF (negative values) channels \cite{MatsEliAngel04}. Thus, ``sparsification'' reduces possible interferences between residual inhibition from past events and the excitation from the present input.\\
Inspired by the locust visual system, a fairly popular class of (SOC-based) computational models and algorithms have been proposed and are under ongoing development (e.g. \cite{RindBramwell96,BlaRinVer99,MatsAngel03gc,MatsEliAngel04,GustavoElAl05,YueEtAl06,FuHuPe18,CizekFaigl19,FuHuPe20,LePeLi23}). The results of three instances of this class of models will be used as reference to compare them with the proposal of this paper.\\
The model proposed in this paper takes a rather unusual approach, using a modern Hopfield network, whose output correlates with the angular size of an approaching object.  Our focus is on background suppression.  Numerical experiments suggest that Hopfield-based collision signaling has a comparable performance to (SOC-based) LGMD models.   In particular, the Hopfield model can signal an object approach when plain SOC does not show a corresponding increase in activity at the end of an approach. This implies that a SOC-based LGMD model would miss the approaching object. However, for certain video sequences the Hopfield model is outperformed by the LGMD models.  These limitations are are linked to the specific nature of information processing in both models and will be described further down.
%
%
%--------------------------------------------------------------
\section{Material and Methods\label{SecMethods}}
%--------------------------------------------------------------
% .................................................................................
\subsection{Hopfield Networks\label{SecHopfieldOverview}}
% .................................................................................
%
\def\softmax{\mathit{softmax}}

\def\pat{x}
\def\vecpat{\vec{\pat}}

\def\query{\xi}
\def\vecquery{\vec{\query}}

\def\prob{p}
\def\vecprob{\vec{\prob}}

\def\key{\mathbf{X}}	% pattern matrix (key)
\def\val{\mathbf{Y}}	% value natrix
Throughout the paper, capital letters denote matrices, and lower case letters represent column vectors.  Vectors are denoted either by $\vec{v}_k$ (where the index $k$ is a label), or as $v_i$ (where the index refers to the element of the $i$-th row of $\vec{v}$).  In order to denote the $i$-th element of $\vec{v}_k$, we use $[\vec{v}_k]_i$.\\
A Hopfield network is an auto-associative memory where the stored pattern vectors $\vecpat_i$ constitute attractors provided that a Lyapunov (or energy) function exists.  Thus, if the network is set to some initial state $\vecquery(t=0)$, it will evolve to a (local) minimum of the Lyapunov function. The originally proposed Hopfield network admits only binary states and pattern \cite{Hopfield82,Hopfield84}, and has a comparatively low storage capacity.  Correlations between stored pattern reduce storage capacity further and typically generate retrieval states which are  linear combinations of the correlated pattern.
\footnote{Some illustrations for the classical Hopfield network with input versus retrieved pattern can be found via the following URLs:
\emph{(i)} \href{https://youtu.be/P6AI1EU9_Uo?si=-h9-C3Y0vB3zrVDH}{adding noise to the input};
\emph{(ii)} \href{https://youtu.be/jYyAy_AkISU?si=OzCtdFLw8OKIlGYX}{rotating the input};
\emph{(iii)} \href{https://youtu.be/gI522seAQlk?si=GnLmffDKGtaUB6Ru}{using contours}.
For all illustrations, the query pattern was stored in the Hopefield network.}\\
Dense associative memories (aka modern Hopfield networks) extend classical Hopfield networks such that storage capacity grows super-linearly \cite{KrotovHopfield2016} or even exponentially \cite{DemircigilEtAl2017} with the number of network units.  This is achieved by using nonlinear functions with a narrow support around the stored pattern, leading to smaller basins of attraction.\\
The binary pattern restriction has been overcome recently \cite{RamsauerEtAl2020, WidrichEtAl2020, KrotovHopfield2020}, while maintaining exponential storage capacity and one (or few) update steps until convergence
\footnote{Illustrations of input versus retrieved pattern using a modern associative memory:
\emph{(i)} \href{https://youtu.be/osSdC5IVfiw?si=LjxCvayrLLInSyaG}{adding noise to the input};
\emph{(ii)} \href{https://youtu.be/g4yNkkvMDzQ?si=PByGMYXMPFdRus6F}{rotation (gray-scale)};
\emph{(iii)} \href{https://youtu.be/sJZfzKhZQbY?si=KS7RkG0bQKyL3Lun}{rotation (binary input)};
\emph{(iv)} \href{https://youtu.be/_nPUWI1LHl4?si=u_-vrIVuPsdoE8u9}{using contour images}.
For all illustrations, the query pattern was stored in the pattern memory.}.
Accordingly, one update rule is based on the $\softmax$-function, which essentially corresponds to the attention mechanism of transformer networks \cite{VaShPa2017},
\begin{equation}\label{EqAttention}
\vecquery(t+1) =\val \cdot \softmax(\beta\, \key^T \cdot \vecquery(t))
\end{equation}
The continuous state of the network ("query") is $\vecquery(t)$; $\beta$ is the inverse temperature. The continuous-valued $d\times 1$ pattern vectors $\vecpat_i \in [-1,+1]$, $1 \leq i \leq N$  are stored as columns of the pattern matrix $\key=\left[\vecpat_i, \vecpat_2 , \ldots, \vecpat_N\right]$.  For an auto-associative memory, $\val \equiv \key$ (in transformers, $\val$ is the value matrix).\\
The $\softmax$-function is defined as
\begin{equation}\label{EqSoftmax}
\softmax(q_i) = \frac{\exp(q_i)}{\sum_j \exp(q_j)}
\end{equation}
The argument of $\softmax(\cdot)$ of Equation \eqref{EqAttention} computes the inner product of the state $\vecquery$ with all stored pattern $\vecpat_i$.  Assuming proper pattern normalization, the resulting vector can be interpreted as the initial probability distribution across the stored pattern $\key$.  If all pattern in $\key$ are well separated (i.e., no two pattern are similar to each other)\footnote{By defining \emph{separation} as $\Delta_i \equiv \min_{1\leq j \leq N, j\neq i}(\vecpat_i^T \vecpat_i-\vecpat_i^T\vecpat_j)$, Theorem 5 in ref. \cite{RamsauerEtAl2020} states that the retrieval error for pattern $\vecpat_i$ decreases exponentially with $\Delta_i$.}, then the pattern with the highest probability typically is selected after one iteration of Equation \eqref{EqAttention}, while all others are suppressed. However, correlations among some of the stored patterns can result in the retrieval of meta-stable states \cite{RamsauerEtAl2020}: Instead of one retrieved pattern, a mixture of similar pattern may appear, because more than one element of $\vecprob(t)\equiv\softmax(\beta\, \key^T \cdot \vecquery(t))$ is close or equal to the maximum.  This can be mitigated by setting $\beta$ to a bigger value, although this may result in numerical issues.\\
%
% .................................................................................
\subsection{A Modern Hopfield Network for Collision Detection\label{SecHopfieldModel}}
% .................................................................................
%
%
\def\roi{\mathbf{R}_{oi}}
\def\Sobel{\mathbf{S}_{odd}}
\def\frame{\tilde{\mathbf{F}}}
\def\pframe{\mathbf{F}}
%
% centering is necessary for stacking the figure panels
\begin{figure}[t!]
 \centering
  \includegraphics[width=0.6\linewidth]{\pics 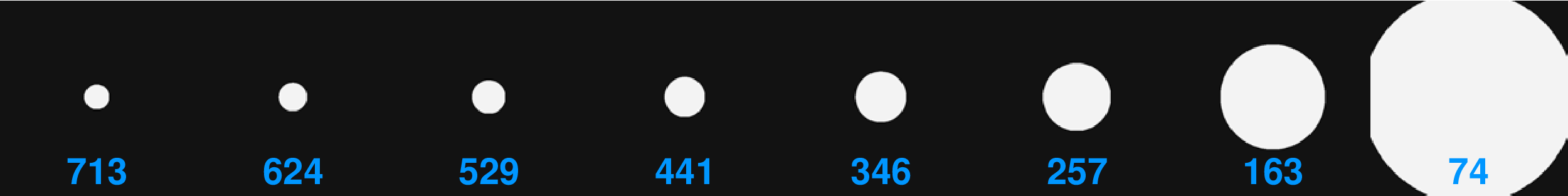}
  \includegraphics[width=0.75\linewidth]{\pics 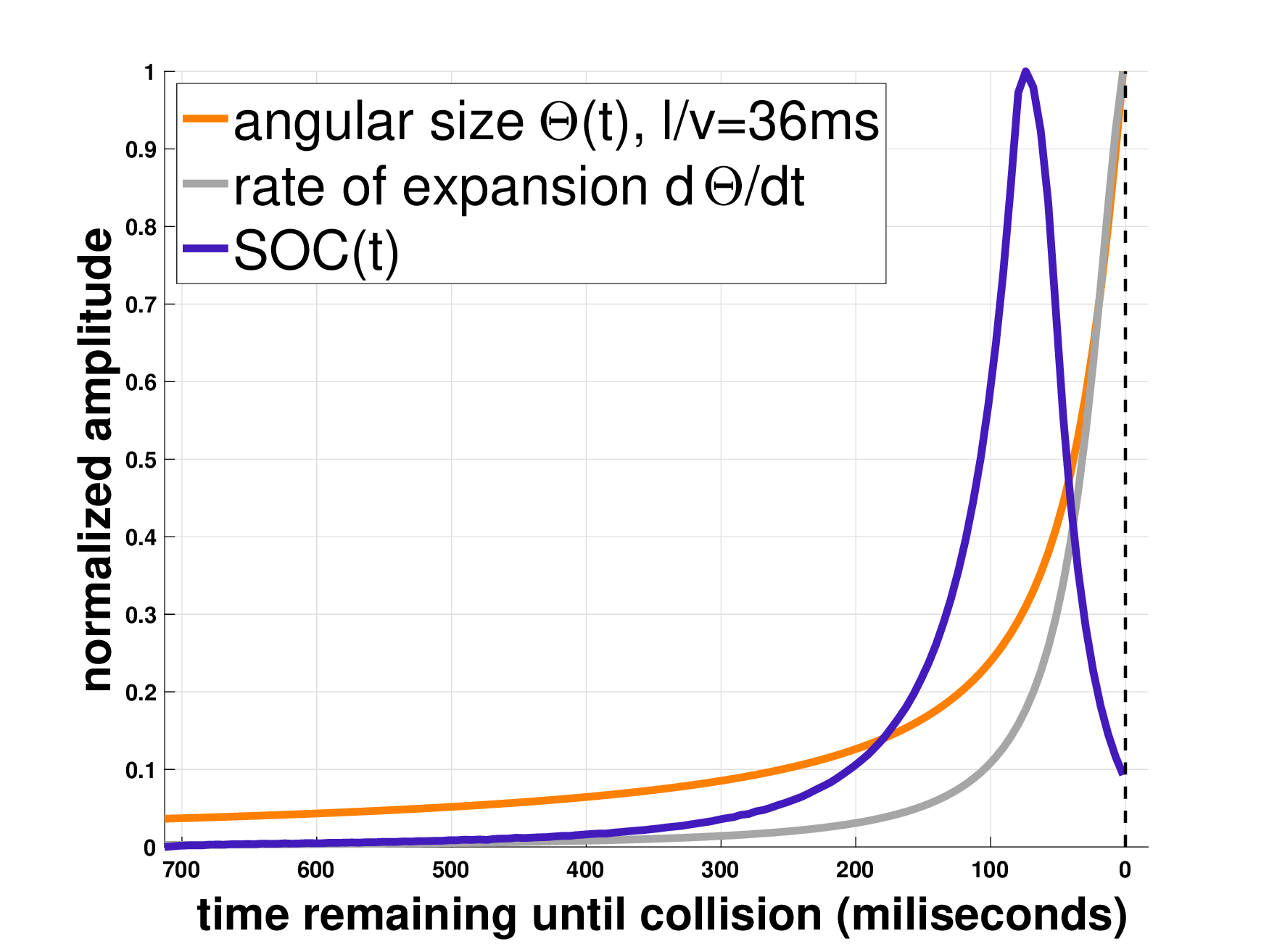}
\caption{\label{FigApproachCurves}{\bf Vanilla Approach Example} Angular size $\Theta(t)\equiv 2\arctan(l/s(t))$, angular velocity $d\Theta/dt=2lv/(s^2(t)-l^2)$ (rate of expansion), and $s(t)=v(\mathrm{ttc}-t)$ with $\mathrm{ttc}=$ time-to-contact (or time-to-collision \cite{MatsJoan2012}).  The sum-of-temporal contrast (SOC$=\sum_{ij}|\frame_{ij}(t)-\frame_{ij}(t- dt)|$, $dt=5.56ms$, indices $i,j$ denote video frame positions) of an approaching uniform disk with half-size-to-velocity ratio $l/|v|=36ms$.  This conforms, for example, to a simulated disk diameter $2l=1m$ and approach velocity $v=50km/h$.  The SOC curve shows a maximum before \textit{ttc}, because for $t\leq 74ms$ the texture-less disk exceeds the boundaries of the image frame causing temporal contrast to decrease.  The top panel shows some of the video frames $\frame_{ij}(t)$ of the disk at the indicated times $t$ in milliseconds.  Ideally, one of these curves should be reproduced by a model which signals object approaches: \emph{(i)} there should be no activity at the beginning, \emph{(ii)} activity should increase super-linearly in the final approach phase, and \emph{(iii)} the curves should be smooth (i.e., noise-free). In the presence of background movements, however, these three characteristics can be severely compromised. %Background movement, however, may severely derogate these three characteristics.
}  
\end{figure}
In this section, an algorithm (computational model) for detecting objects that approach the observer on a direct collision course is defined. We emphasize reproducibility by providing step-by-step instructions. The selected parameter values were determined through a systematic exploration of the parameter space with the goal of achieving ``good'' overall performance with a set of benchmark videos.  Good performance means that the model's response follows the angular size of the approaching object (Figure \ref{FigApproachCurves}): when the object is far away, the response should be small or zero, and when the object is close, the response should increase almost exponentially, while being as smooth as possible.  Specifically, the benchmark set consisted of four artificial and four real-world video sequences.

%
% .................................................................................
\subsubsection{Video Frame Processing\label{SecFrameProcessing}}
% .................................................................................
%
%
% centering is necessary for stacking the figure panels
%
\begin{figure}[t!]
 \centering
 \subfloat[video frame $\frame(t)$]{\includegraphics[width=0.23\linewidth]{\pics 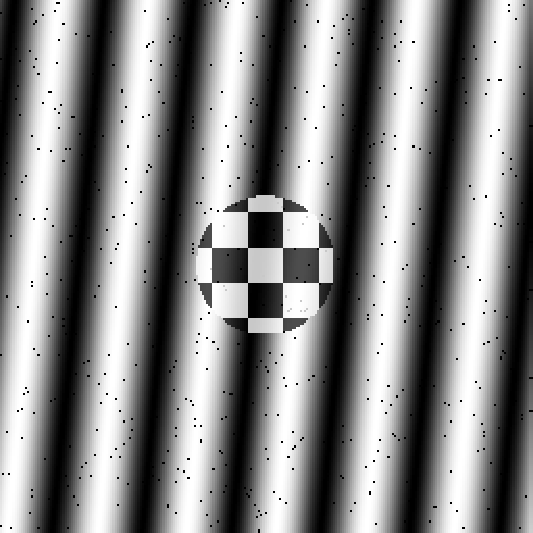}}~
 \subfloat[$\frame(t) \ast \Sobel$]{\includegraphics[width=0.23\linewidth]{\pics 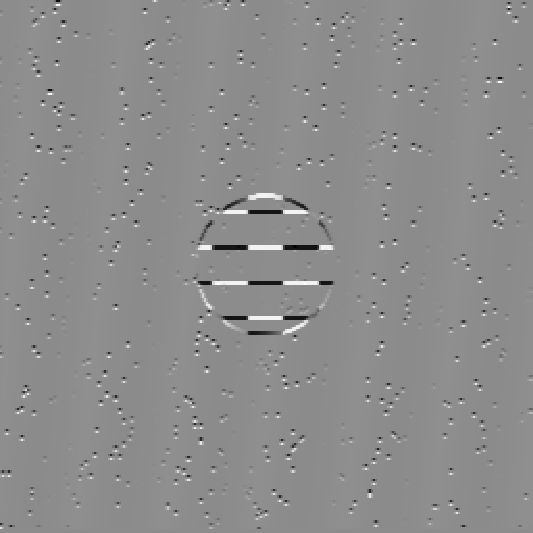}}~
 \subfloat[mask $\roi$]{\includegraphics[width=0.23\linewidth]{\pics 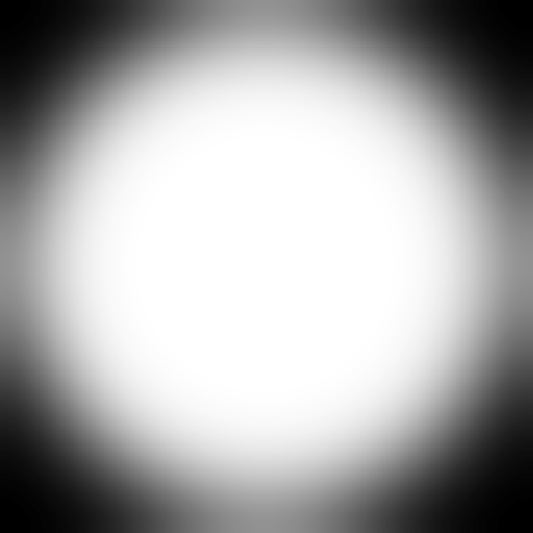}}~
 \subfloat[$\roi \odot (\frame(t) \ast \Sobel)$ ]{\includegraphics[width=0.23\linewidth]{\pics 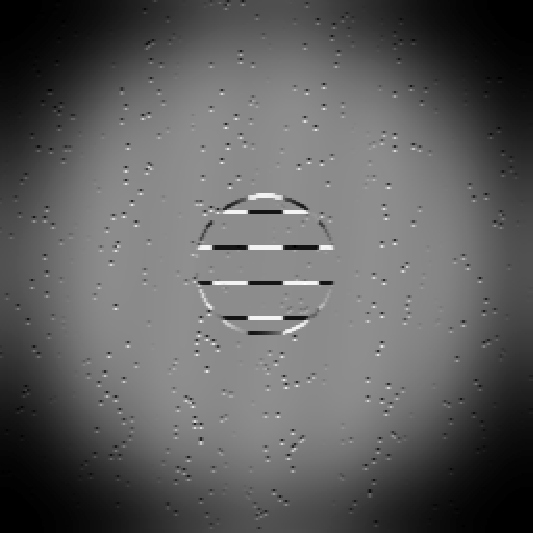}}
\caption{\label{FigVideoProcessing}{\bf Video Processing Pipeline} \textbf{(a)} The input is a gray-level video frame $\frame(t)$
\textbf{(b)} $\frame(t)$ convolved with Sobel-Feldman operator $\Sobel$ (Equation \eqref{EqSobelFeldmanKernel})
\textbf{(c)} Region-of-interest $\roi$
\textbf{(d)} Finally processed video frame $\pframe(t)$ (Equation \eqref{EqFrameProcessing}).
}
\end{figure}

Let $\frame_{ij}(t) \in [0,1]$ be a gray-level video frame with $r$ rows and $c$ columns at discrete time $t=1,2,3,\ldots,t_{max}$.  Video frames with $r \neq c$ can be symmetrically embedded in a square matrix with a constant luminance (e.g. $0.5$) such that the resulting frame has an equal number $n=\max(r,c)$ of rows and columns, respectively.  Proceeding so, however, may compromise the detection of potential collisions.  Alternatively, the frames can be cropped such that $n=\min(r,c)$.\\

Let $\Sobel$ be a convolution kernel implementing a modified version of the Sobel-Feldman operator (cf. \cite{Scharr00}) for enhancing the horizontal edges of $\frame(t)$ (Figure \ref{FigVideoProcessing}b),
\begin{equation}\label{EqSobelFeldmanKernel}
\Sobel=\frac{1}{16}
\begin{bmatrix}
 3 &  10 &  3\\
 0 &   0 &  0\\
-3 & -10 & -3\\
\end{bmatrix}
\end{equation}
A region-of-interest (mask) $\roi$ is constructed by \emph{(i)} creating a white disk (luminance $1$) on black background (luminance $0$) with radius $\rho=0.9n/2$, and \emph{(ii)} blurring the disk with a Gaussian kernel with standard deviation $\sigma=20$ pixel (Figure \ref{FigVideoProcessing}c).  Applying the mask to the video frames improves specifically performance for videos with strong background motion.  Videos without background motion would not benefit much from the mask.  Neither the value of $\rho$ nor the degree of spatial blur $\sigma$ turned out to be critical for the considered video footage.  The value of $\rho$ was selected from  $2\rho/n= \in\{ 0.5, 0.75, 0.9\}$ plus ``\emph{no mask at all}''.  The degree of spatial blur was selected from $\sigma \in \{ 0.1, 10, 20, 40 \}$.\\
Figure \ref{FigVideoProcessing}d shows the final result $\pframe(t)$ of video processing: At each time $t$, frame $\frame(t)$ is first convolved with $\Sobel$ (symbol "$\ast$") and then element-wise multiplied with $\roi$ (Hadamard product, symbol "$\odot$"),
\begin{equation}\label{EqFrameProcessing}
\pframe(t) = \roi \odot \left[  \frame(t) \ast \Sobel  \right]
\end{equation}
%
% .................................................................................
\subsubsection{Image to Vector Conversion\label{SecImg2vec}}
% .................................................................................
%
\def\template{\tilde{\mathbf{P}}}
\def\ptemplate{\mathbf{P}}

In order to be used with modern Hopfield networks, $n \times n$ (image) matrices $\mathbf{V}$ ($=$ video frames $\pframe$ and template pattern $\ptemplate$) have to be converted into  $d \times 1$ vectors $\vec{v}$ ($\vecquery$ and $\vecpat_i$, respectively) with $d=n^2$.  All vectors are assumed to have the following properties:
\begin{enumerate}
\item $v_{i+(j-1)n} = \mathbf{V}_{ij} \ \forall i,j=1,2,\ldots,n$ (index mapping)
\item  $\sum_k v_k=0$ (centering at zero)
\item $\sum_k v^2_k=1$ (normalization) 
\end{enumerate}
The first property states that all matrices have to be converted into vectors in the same way (i.e., with identical index mapping).  The second property is implemented by subtracting the mean $\vec{v}\leftarrow \vec{v}-\sum_k v_k/d$.  The third property is implemented by dividing by the Euclidean norm $\vec{v}\leftarrow \vec{v}/\|\vec{v}\|_2$.  Thus, all vectors are unit vectors.

% .................................................................................
\subsubsection{Pattern Memory\label{SecPatternMemory}}
% .................................................................................
%
\def\on{\mathbf{ON}}
\def\off{\mathbf{OFF}}
\begin{figure}[t!]
 \centering
  \includegraphics[width=1.0\linewidth]{\pics 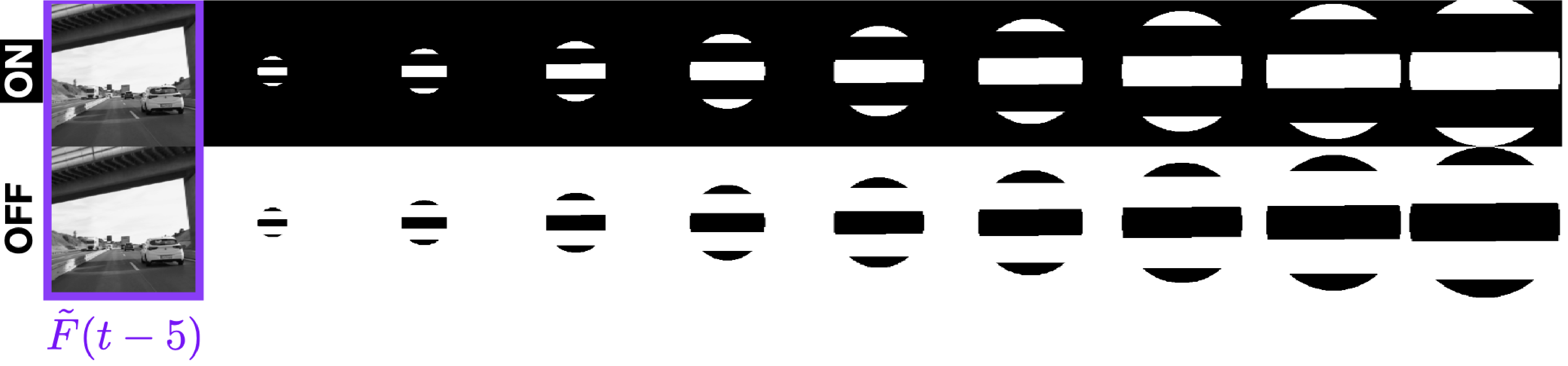}
\caption{\label{FigPatternMemory}{\bf Dynamic Pattern Memory} This is an illustration which shows the unprocessed images. Note that in the actual pattern matrices, vectors corresponding to the \emph{processed} video frame and template pattern are stored in order of columns. The first column contains the delayed and processed video frame, while the second column has the smallest version of the template pattern. The size of the template pattern increases with each subsequent column, such that the last column corresponds to the template pattern with its original size. The template image was selected after trying additionally a homogeneous disk with constant luminance, a noise patch, a checkerboard, and a vertically oriented grating.  The checkerboard, the horizontal and the vertical grating were each tried with spatial frequencies of $2,4,8$ and $16$ cycles per image.
}
\end{figure}
\begin{figure}[t!]
  \includegraphics[width=0.75\linewidth]{\pics 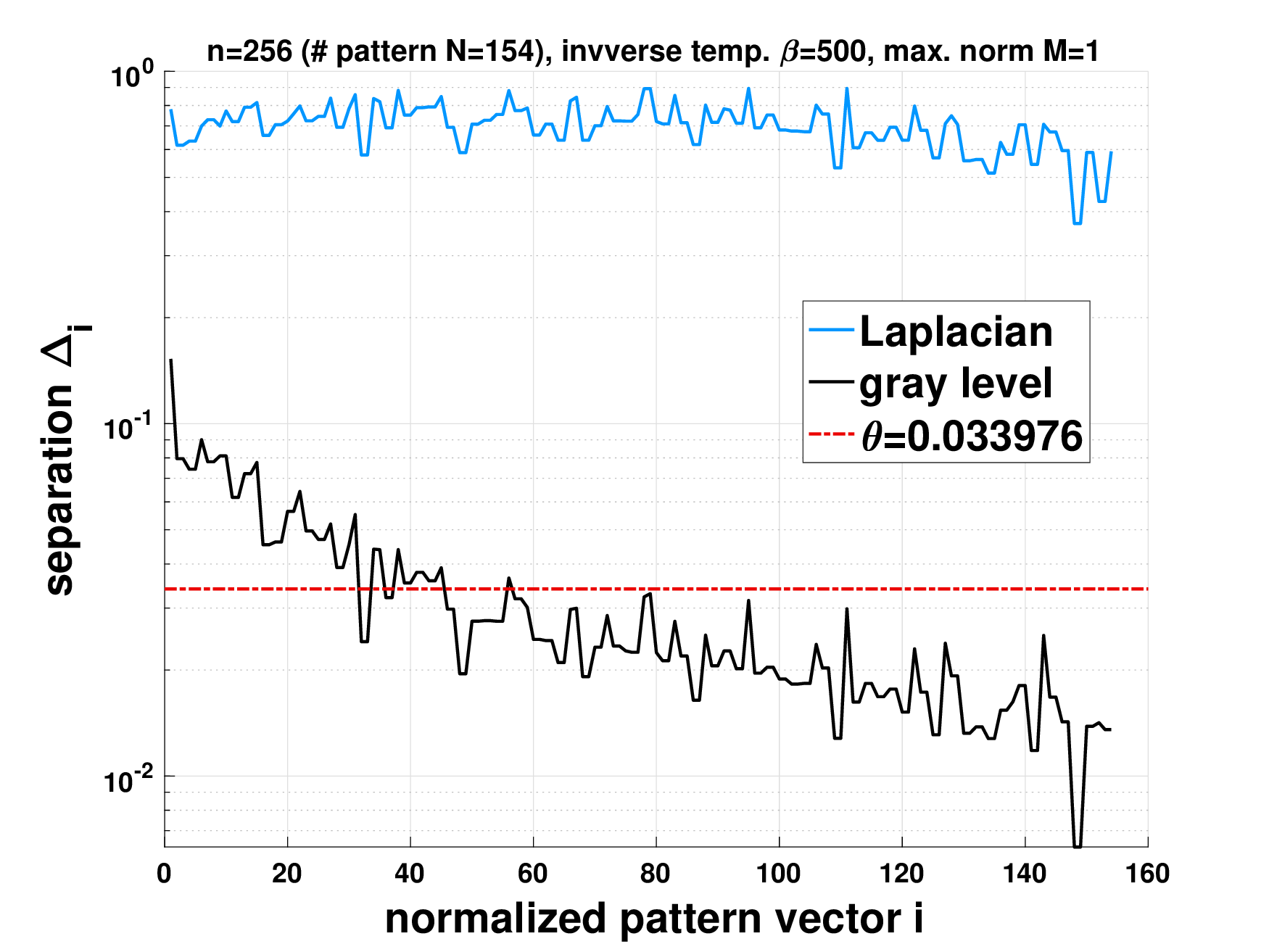}
\caption{\label{FigSeparation}{\bf Pattern Separation} The figure shows the pattern separation $\Delta_i \equiv \min_{1\leq j \leq N, j\neq i}(\vecpat_i^T \vecpat_i-\vecpat_i^T\vecpat_j)$ according to ref. \cite{RamsauerEtAl2020}. The pattern memory consisted of $N=154$ pattern vectors $\vecpat_i$.  The dynamically updated \& delayed video frame was omitted (thus for this plot $i=1,2,\ldots,N$ with $N=1+\lfloor3n/5\rfloor$).  The pattern templates ($\template(s)$ and $\ptemplate(s)$, respectively) had $n=256$ rows and columns (i.e., $d=65536$, as shown in Figure \ref{FigPatternMemory}).  Scaling $s$ increases with $i$ (abscissa).  In the plot, two instances of pattern memories are compared.  First, one generated from the originally gray-level template $\template(s)$  (legend label \emph{gray-level}).  Second, one using their highpass filtered versions $\ptemplate(s)$  (label \emph{Laplacian}, cf. Equation \eqref{EqLaplacian}).  The horizontal line denotes the threshold $\Theta \equiv 2/(\beta N) + [\log(2(N-1)N\beta M^2)]/\beta$ \cite{RamsauerEtAl2020}.  If the pattern $\vecpat_i$ is well separated, then $\Delta_i \geq \Theta$.  Whereas the separation of the \emph{gray-level} pattern decreases with $s$ (and thus with disk diameter), their highpass filtered versions \emph{Laplacian} are largely independent of $s$.  This is because luminance images have more spatial redundancy than images with enhanced contrast boundaries.}
\end{figure}
Pattern vectors $\vecpat_i$ are columns of $d \times (N+1)$ pattern matrices (pattern memory $\key$, cf. Equation \eqref{EqAttention}).  Specifically we have two separate pattern memories
\begin{eqnarray}\label{EqOnOffMemory}
\on & \equiv & \left[\vecquery(t-5), \vecpat_i,  \vecpat_2 ,  \ldots,  \vecpat_N\right]\\
\off &\equiv & \left[\vecquery(t-5), -\vecpat_i, -\vecpat_2 ,  \ldots, -\vecpat_N\right]
\end{eqnarray}
This implies two corresponding instances of Equation \eqref{EqAttention}, one with $\val=\key=\on$, and another one with $\val=\key=\off$.\\
Let $\vecquery(t)$ be the pattern vector from converting the processed video frame $\pframe(t)$ with the properties defined in Section \ref{SecImg2vec}.  Then,  $\on_1=\off_1=\vecquery(t-5)$.  That is, the first column of both pattern memories is dynamically updated with a by five time steps delayed video frame (Figure \ref{FigPatternMemory})\footnote{For initialization, the video sequence may be started at $t=6$, or $\vecquery(t-5)=\vecquery(t)$ for $t<6$.  The delay by five time steps was selected from $5$, $10$, and $15$.  We tried furthermore an adaptive delay, but discarded it because it did not lead to a significant improvement over the fixed delay.}.  The remaining $N$ columns of $\on$ and $\off$ do not change with time and are laid out as follows.\\
Let $\template_{ij}(s) \in [0,1]$ be a gray-level template image with $n$ rows and columns, respectively.  The default template image is a disk with an overlaid horizontal grating (2 cycles per image) as shown in Figure \ref{FigPatternMemory}.  Notice that the grating orientation matches the orientation of the Sobel-Feldman operator for processing the video frames (Equation \eqref{EqSobelFeldmanKernel}).\\
In total, $1+\lfloor3n/5\rfloor$ versions of the template image $\template (s)$ are generated, with varying disk diameters of $s\cdot n$ pixel, starting with a scale-factor of $s=0.1$ and increasing to $1$ in steps of $\Delta s=3/2n$\footnote{This heuristics for $s$ has been found by trying starting values $s \in \{0.1, 0.25, 0.5\}$ combined with increments $\Delta s \in \{3/2n, 3/n, 6/n\}$.  Results are not significantly different for most videos if instead we start with a constant disk diameter of seven pixel $s=7/n$ and use increments of $\Delta s=1/n$.}.
Subsequently, the edges of each template image are enhanced with the Laplacian operator,
\begin{equation}\label{EqLaplacian}
\ptemplate(s) \equiv \nabla^2 \, \template(s)
\end{equation}
and converted into vectors $\vecpat_i$ with the properties defined in Section \ref{SecImg2vec}.  The $\vecpat_i$ are stored in $\on$ from column $2$ to $N$.  The inverse versions $-\vecpat_s$ are stored in $\off$.  Therefore, the size of $\on$ and $\off$ is $d \times N$, with $d=n^2$ and $N=2+\lfloor3n/5\rfloor$ (one more because of $\vecquery(t-5)$).  Compared with storing gray-level images, their edge-enhanced counterparts are better separated (Figure \ref{FigSeparation}).  Notice that the pattern matrices have to be specifically build according to video frame size $n \times n$.\\
Several combinations of functions operating on the pixels of the video frames were tried for Equations \ref{EqSobelFeldmanKernel} and \ref{EqLaplacian}, respectively:
\begin{enumerate}
\item Luminance (unprocessed video frames)
\item Laplacian
\item Difference of successive video frames (temporal contrast)
\item Difference between the gradient magnitudes of successive video frames
\item Odd-symmetric Sobel-Feldman operator with $8$ orientations
\item Even-symmetric Sobel-Feldman operator ($5 \times 5$ LDL kernel, where ``L'' (light) means positive values of the filter kernel and ``D'' (dark) stands for negative values) with $4$ orientations
\item Oriented temporal contrast along $4$ directions using the following calculations: \textit{(i)} Temporal contrast; \textit{(ii)} convolution of absolute temporal contrast with an oriented and even-symmetric $5 \times 5$  Sobel-Feldman kernel; \textit{(iii)} spatial blurring the result of the previous step with an asymmetric Gaussian kernel (standard deviation $10$ and $0.1$, respectively) with perpendicular orientation \textit{(iv)} multiplication with temporal contrast.
\end{enumerate}
When oriented operators were used for both video frame processing and pattern memory processing, then their respective orientations were matched .  Furthermore, ``temporal contrast'' for the pattern memory is defined as the difference $\template_{ij}(s+\Delta s) - \template_{ij}(s)$.
%
%
% .................................................................................
\subsubsection{Pattern Retrieval\label{SecPatternRetrieval}}
% .................................................................................
%
\def\qon{\query^\mathrm{on}}
\def\qoff{\query^\mathrm{off}}
\def\vecqON{\vecquery^\mathrm{\,on}}
\def\vecqOFF{\vecquery^\mathrm{\,off}}

\def\pon{\prob^\mathrm{on}}
\def\vecpON{\vec{\prob}^\mathrm{\ on}}
\def\poff{\prob^\mathrm{off}}
\def\vecpOFF{\vec{\prob}^\mathrm{\ off}}
The proposed algorithm proceeds frame-wise where $t$ denotes the current video frame.  The current video frame $\frame(t)$ is processed and converted into $\vecquery(t)$.  The queries are initialized with $\vecqON(\tau=0)=\vecquery(t)$ and $\vecqOFF(\tau=0)=\vecquery(t)$, where $\tau$ denotes the iteration number of the update rules
\begin{eqnarray}\label{EqUpdate}
\vecqON(\tau+1) & = & \on \cdot  \vecpON(\tau)\\\nonumber
\vecqOFF(\tau+1) & = & \off \cdot \vecpOFF(\tau)
\end{eqnarray}
with the probability distributions $\vecpON$ and $\vecpOFF$, respectively, defined as
\begin{eqnarray}\label{EqProbability}
\vecpON(\tau) & \equiv & \softmax(\beta\, \on^T \cdot \vecqON(\tau))\\\nonumber
\vecpOFF(\tau) & \equiv& \softmax(\beta\, \off^T \cdot \vecqON(\tau))
\end{eqnarray}
The inverse temperature is set to $\beta=500$, and Equation \eqref{EqUpdate} is iterated until $\|\vecqON(\tau+1)-\vecqON(\tau)\|_2\leq 0.01$ or $\tau\geq 5$ (analogous for $\vecqOFF$).  As mentioned in Section \ref{SecHopfieldOverview}, the update usually converges after one iteration for well separated pattern (Figure \ref{FigSeparation}).\\
The inverse temperature cannot be increased much further for an improved suppression of meta-stable states. The reason is that numerical overflow due to exponentiation may occur\footnote{If numerical problems occur, then one could replace Equation \eqref{EqSoftmax} by a "fail-safe" version $\softmax(q_i) = \exp(q_i)/(\max_k\{q_k\} + \sum_j \exp(q_j))$ along with $\beta=50$.  Doing so would also suppress meta-stable retrieval states.}.\\
%
% .................................................................................
\subsubsection{Conversion of Retrieval States into Neuronal Activity\label{SecRetrieval2activity}}
% .................................................................................
%
\def\activity{z}
\def\veccol{\vec{c}}
\def\hatacton{\hat{\activity}^\mathrm{\,on}}
\def\hatactoff{\hat{\activity}^\mathrm{\,off}}
\def\acton{\activity^\mathrm{on}}
\def\actoff{\activity^\mathrm{off}}

The retrieved pattern are not used any further (except for illustration).  The label $i$ of the winning pattern (or their mean in case of a meta-stable retrieval state) is directly taken as activity.  Let $\veccol\equiv[1,2,\ldots,N]$ be a row vector denoting the column number of pattern memories $\on$ and $\off$, respectively.  The (scalar) activities $\hatacton$ and $\hatactoff$ are defined as the inner products
\begin{eqnarray}\label{EqActivity}
\hatacton  & \equiv & \veccol \cdot \vecpON(\tau) \\\nonumber
\hatactoff  & \equiv & \veccol \cdot \vecpOFF(\tau)
\end{eqnarray}
The higher the activity, the greater the disk diameter of the retrieved pattern.  Therefore, activities are proportional to the angular size of an approaching object.  The lowest activity is obtained if the delayed video frame is retrieved.
Since the activity is usually very spiky, online smoothing (lowpass filtering) is applied,
\def\fin{\mathit{in}}
\def\fout{\mathit{out}}
\def\lp{\mathcal{F}}
\begin{eqnarray}\label{EqLowpass}
 \fout& = &\lp[\fin,\alpha]\\\nonumber
 :\Leftrightarrow  \fout(t+1)&=&\alpha\, \fout(t) + (1-\alpha) \fin(t)
\end{eqnarray}
This equation is a discrete representation of a leaky integrator neuron (cf. \href{https://doi.org/10.1371/journal.pcbi.1002625.s008}{Text S8} of \cite{MatsJoan2012}).  The memory coefficient $0<\alpha<1$ is a constant that sets the degree of smoothing; $\fout$ is the state variable and the smoothed output; $\fin$ is the original signal (filter input).  If $\alpha$ is close to one (equivalent to a small leakage conductance), then the input is mainly integrated and thus strongly smoothed.  However, more smoothing translates into a greater delay (i.e., phase lag) between input and output.  This has to be taken into account for real-time applications of the proposed algorithm.  Accordingly,
\begin{eqnarray}\label{EqLPActivity}
\acton  & = & \lp[ \hatacton,\alpha]\\\nonumber
\actoff & = & \lp[ \hatactoff,\alpha]
\end{eqnarray}
where  $\alpha=0.85$ and $\acton(t)$ \& $\actoff(t)$ are the smoothed activities ($=$ filter output). Finally, the ON and OFF signals are combined by multiplication \cite{CarpenterGrossberg81,GabKraKocLau02,NIPS2011_0348,Mats2015},
\begin{equation}\label{EqMultiplication}
\activity(t) \equiv \acton(t) \cdot \actoff(t)
\end{equation}
Notice that $\min_t \activity(t)=1$ because $\acton(t),\  \actoff(t) \in [1,N]$.  This is to say that because the minimum activity of either channel is nonzero, $\activity(t)$ will reflect the activity of at least $\acton(t)$ or $\actoff(t)$.
% 
% .................................................................................
\subsection{Benchmark Models\label{SecBenchmarkModels}}
% .................................................................................
%
\def\hopfieldx{\emph{``Hopfield''}}
\def\advancedx{\emph{``Advanced''}}
\def\hexx{\emph{``CizFai19''}}
\def\yuex{\emph{``FuHuPe20''}}
\def\hopfield{{\hopfieldx} }
\def\advanced{{\advancedx} }
\def\hex{{\hexx} }
\def\yue{{\yuex} }

The results of the \hopfield  model as introduced above will be compared to three models which are based on temporal contrast extraction ("SOC-based models").  These are briefly outlined in what follows.
%
% .................................................................................
\subsubsection{A neural model of the locust visual system for detection of object approaches with real-world scenes \cite{MatsEliAngel04} (\advancedx) \label{MatsEliAngel04}}
% .................................................................................
%
\def\rect#1{[#1]^+}

The model \advanced splits temporal contrast into parallel ON and OFF pathways.  The ON channel encodes luminance increments from one video frame to the next, the OFF channel decreasing luminance.  Lateral inhibition is implemented by self-limiting diffusion layers.  Diffusion is self-limiting because it curtails the excitatory activity by which it is fed.  Let $\acton(t)$ be the halfwave rectified activity of the ON-channel, and $\actoff$ that of the OFF-channel.  The channels are combined by
\begin{equation}\label{EqSigmaPi}
\activity(t) \equiv \lp[ \acton \cdot \actoff + \epsilon ( \acton + \actoff ),\alpha]
\end{equation} 
where $\epsilon=0.001$ and filter memory $\alpha=0.5$.  Different to the \hopfield model, $\acton(t)$ and $\actoff(t)$ of \advanced can be zero.  The first term thus implements a logical ``\texttt{AND}'' gate, which would be zero for example for a white disk approaching against a black background, or an approaching bird against a clear sky.  In order to obtain non-zero activity in the latter cases, the second term was included.
%
% .................................................................................
\subsubsection{Self-Supervised Learning of the Biologically-Inspired Obstacle Avoidance of Hexapod Walking Robot \cite{CizekFaigl19} (\hexx) \label{CizekFaigl19}}
% .................................................................................
%
The \hexx-model computes LGMD1 responses to both lighter and darker objects than the background \cite{RindSimmons92a}.  It was implemented following the equations seven (``photoreceptor layer'', i.e. temporal contrast) to eleven (``summation layer'') of Section 3.3. in ref. \cite{CizekFaigl19}.  Equation twelve eventually sets all units $S_{ij}$ of the summation layer to zero if $S_{ij}\leq T_s$.  Since the threshold (originally a scalar constant) ``\emph{$T_s$ has to be set experimentally in such a way to avoid saturation of the LGMD output}'', here I replaced it by an adaptive mechanism controlled by temporal contrast activity\footnote{The model is invariant with respect to the range of luminance values.} $P_{ij}(t)\equiv\frame(t) - \frame(t-1)$,
\begin{equation}\label{EqCizekFaigl19a}
T_s  =  \lp\left[\frac{1}{n^2}\sum_{i,j=1}^n |P_{ij}(t)|,  \alpha_s\right]
\end{equation}
where $T_s=T_s(t)$ is scalar, the size of the gray-level video frame $\frame \in [0,1]$ is $n \times n$ and the filter memory coefficient is set to $\alpha_s=0.75$ (cf. Equation \eqref{EqLowpass} above).  Subsequently, all summation layer units $S_{ij}$ with $S_{ij}\leq 2T_s(t)$ are set to zero.  For computing the model's output  $\activity(t)$ (i.e., the LGMD membrane potential), let $\hat{\activity}(t)\equiv\sum_{i,j=1}^n |S_{ij}(t)|/n^2$.  Then:
\begin{equation}\label{EqCizekFaigl19b}
\activity(t)  = \left\{
   \begin{array}{ll}
      \hat{\activity}(t) & \textrm{if $\hat{\activity}(t)>0.95\, T_l(t)$}\\
      0 & \textrm{otherwise}\\
   \end{array} \right.
\end{equation}
subject to another adaptive threshold $T_l(t)$ that varies according to
\begin{equation}\label{EqCizekFaigl19c}
T_l =  \lp\left[ \hat{\activity}, \alpha_l \right]
\end{equation}
with filter memory coefficient  $\alpha_l=0.75$.
%
% .................................................................................
\subsubsection{A Robust Collision Perception Visual Neural Network With Specific Selectivity to Darker Objects \cite{FuHuPe20} (\yuex)\label{FuHuPe20}}
% .................................................................................
%
%
The \yuex-model computes LGMD2 responses to darker objects than the background \cite{SimmonsRind1997}.
As the previous model, it also uses temporal contrast $P_{ij}(t)\equiv 255\,(\frame(t) - \frame(t-1))$ at its front end.  Since the complementary ON-channel is missing, I created it by using a second instance of the model with inverse temporal contrast $\tilde{P}\equiv-P$ as input.  After conducting tests with a variety of video footage, the (free) frame rate parameter of the model was set constant to $120$ Hz.  The spiking output (i.e., equation 27 in ref. \cite{FuHuPe20}) of both model instances  $\acton(t)$ and $\actoff(t)$, respectively, was combined according to Equation \eqref{EqSigmaPi} with $\epsilon=0.05$ and filter memory $\alpha=0.75$.  Lowpass filtering of the spikes essentially mirrors the membrane potential $\activity(t)$ of a (hypothetical) neuron postsynaptic to the LGMD.
%
%
%--------------------------------------------------------------
\section{Results\label{SecResults}}
%--------------------------------------------------------------
%
%--------------------------------------------------------------
\subsection{Principal Characteristics and Limitations\label{SecProperties}}
%--------------------------------------------------------------
%
\def\fps{\mathrm{fps}}	% frames per second in Hz
\def\cpix{$\mathrm{cyc/img}$}	% cycles per image
\def\cpi{$\mathrm{cyc/img}\ $}	% cycles per image
\def\dptx{$\mathrm{deg/t}$}		% degree per frame number (not used)
\def\dpt{$\mathrm{deg/t}\ $}	% degree per frame number (not used)
\begin{figure}[t!]
 \centering
 \subfloat[drifting horizontal grating ($k_s$ varies)]{\includegraphics[width=0.49\linewidth]{\pics 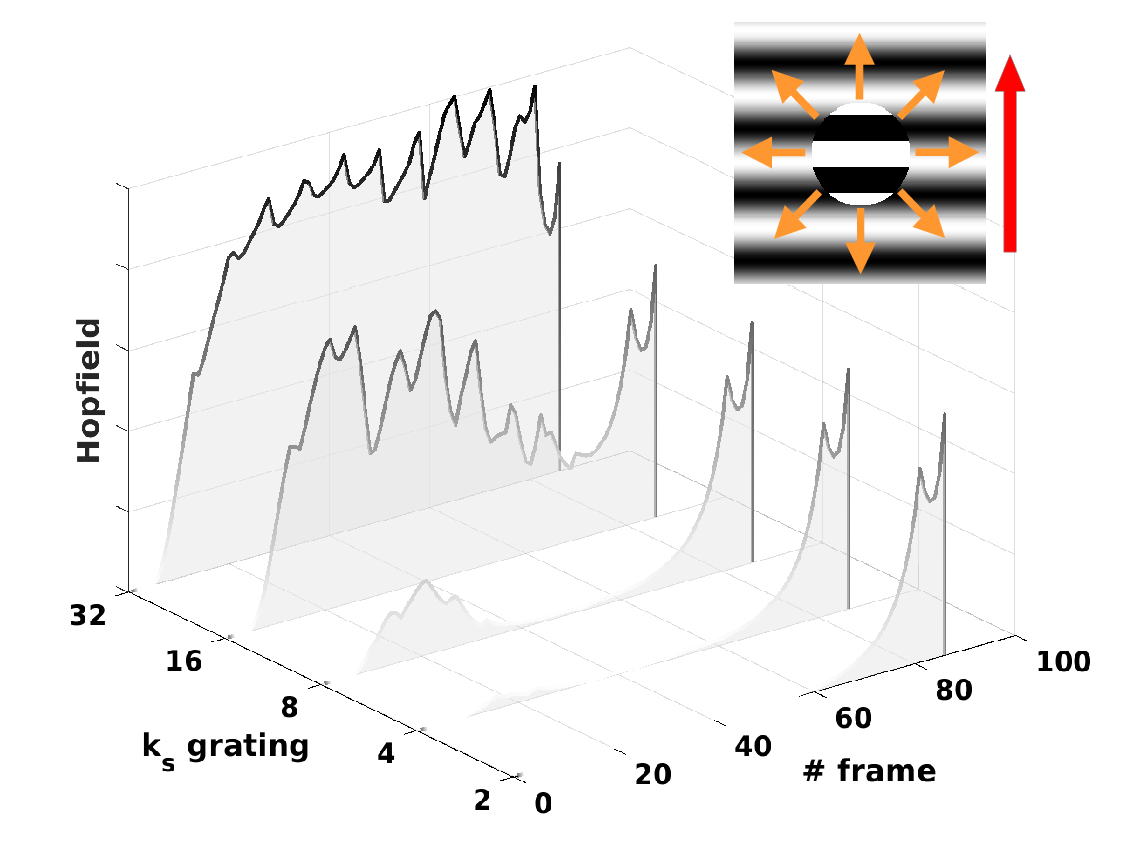}}~
 \subfloat[drifting vertical grating ($k_t$ varies)]{\includegraphics[width=0.51\linewidth]{\pics 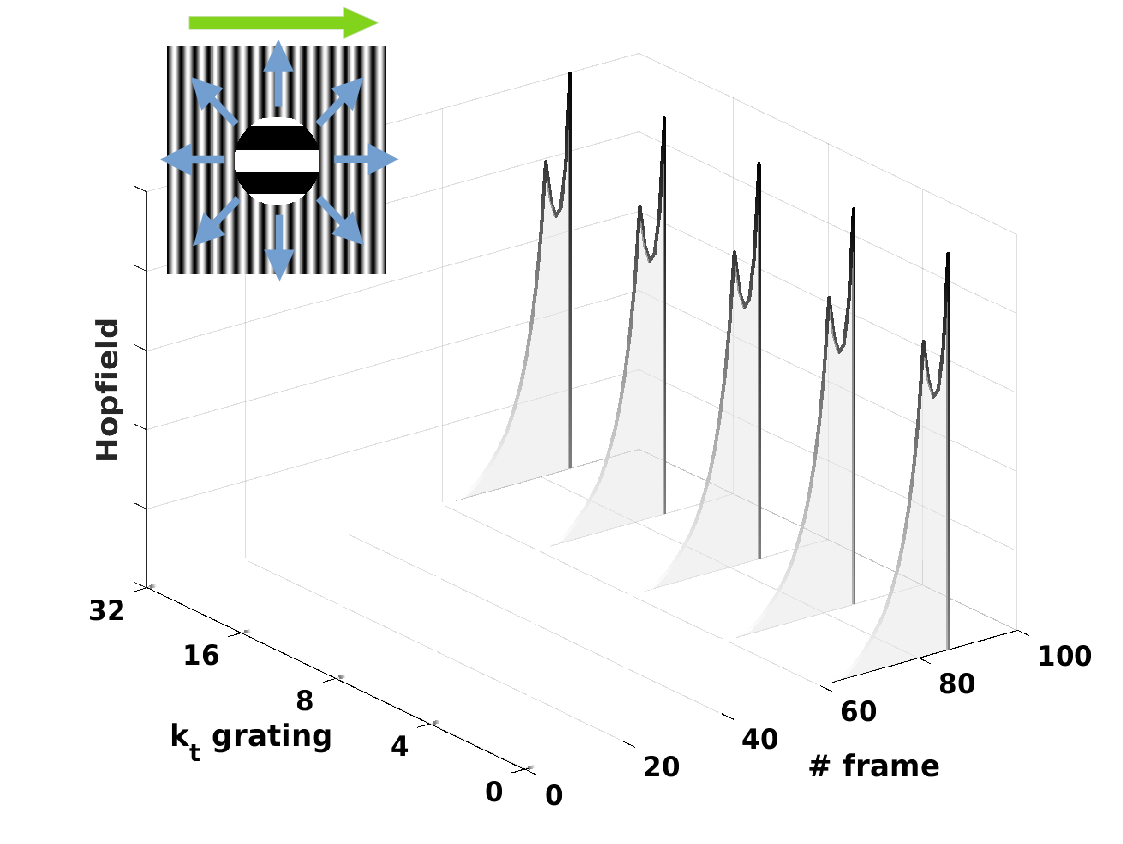}}
\caption{\label{FigGratingOrientation}{\bf Grating orientation}.  Responses of the \hopfieldx-model to an approaching disk over a drifting horizontal and vertical grating, respectively. The drifting direction  of each grating is indicated by an arrow.  The disk's texture is a horizontal square-wave grating of $2$ cycles per disk (insets: frame \#73; frame size $256\times 256$), matching the spatial frequency and orientation of the template pattern (cf. Figure \ref{FigPatternMemory}).  The disk has a diameter of $0.5$ m and approaches with a constant speed of $50$ km/h from an initial distance of $10$ m.  Its final distance from the observer is $0.1$ m.  With a sampling rate of $120$ frames per second, the video comprises $86$ frames.
\textbf{(a)} Results for spatial frequencies $k_s=2,\ 4,\ 8,\ 16\ \mathrm{and}\ 32$ \cpi (i.e., cycles per image) of the grating.  Grating orientation is horizontal, and drifting speed is $k_t=8$ Hz.  The figure reveals the highpass characteristics of \hopfieldx, as higher spatial frequencies of the background grating cause more undesired pattern retrievals.  Undesired pattern retrievals make the curves noisier, and responses to $k_s\geq32$ \cpi do not contain any useful information. 
\textbf{(b)} Grating ($k_s=16$ \cpix ) orientation is vertical and thus does not match that of the template pattern (Figure \ref{FigPatternMemory}).  As a consequence no interference occurs.  Exactly the same plot (i.e., identical results) is obtained for varying $k_s$ as in the left panel, but with a vertical grating (not shown).}
\end{figure}
In this section, we explore some of the main characteristics of the \hopfieldx-model by means of artificial video sequences.  The sequences can be fully parameterized in terms of physical approach variables (cf. Figure \ref{FigApproachCurves}), object type (e.g., spatial frequency and orientation of the rectangular grating that forms the texture of the approaching disk), and background (e.g. a sine wave grating with a certain spatial frequency, orientation, and drifting speed). Drifting speed was implemented as the time dependent phase $2\pi t k_t \Delta t$ of the grating where $t$ is the frame number and $\Delta t=1/\fps$ with $\fps=120$ Hz.\\
The determining factor for constructing a worst case scenario for the \hopfieldx-model is the orientation of the background grating.  When the grating is horizontally oriented (as the template pattern shown in Figure \ref{FigPatternMemory}), then pattern retrieval is compromised  (i.e., interference occurs) for certain combinations of the grating's drifting speed and spatial frequency.  Similarly, variation of the  spatial frequency and orientation of the approaching disk's texture only impairs pattern retrieval if the background grating is horizontally oriented.  In contrast, variations of disk or grating parameters would not interfere with pattern retrieval when the grating is vertically oriented: Corresponding data are only insignificantly different from to those shown in Figure \ref{FigGratingOrientation}b.\\
Assuming a horizontally oriented background grating, the critical parameter is its spatial frequency $k_s$ (Figure \ref{FigGratingOrientation}a): Increasing $k_s$ will increase interference and therefore overall retrieval activity. For $k_s=32$ \cpix, retrieval activity will no longer increase in the final approach phase. These properties of the \hopfieldx-model are linked to high-pass filtering: The \hopfieldx-model predominantly uses the high spatial frequencies of video frames and pattern templates (Equations \eqref{EqFrameProcessing} and \eqref{EqLaplacian}, respectively).\\
\begin{figure}[t!]
 \centering
 \subfloat[SOC ($k_t$ varies)]{\includegraphics[width=0.48\linewidth]{\pics 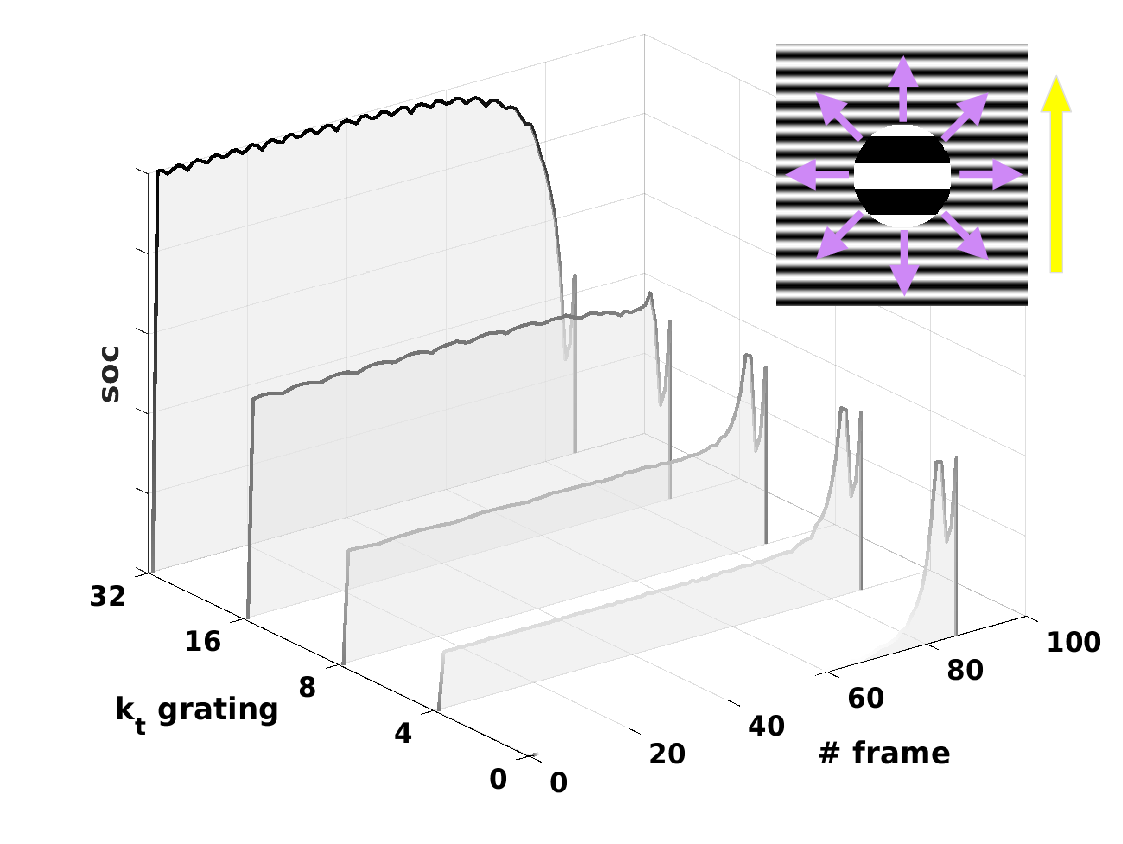}}~
 \subfloat[Hopfield ($k_t$ varies)]{\includegraphics[width=0.52\linewidth]{\pics 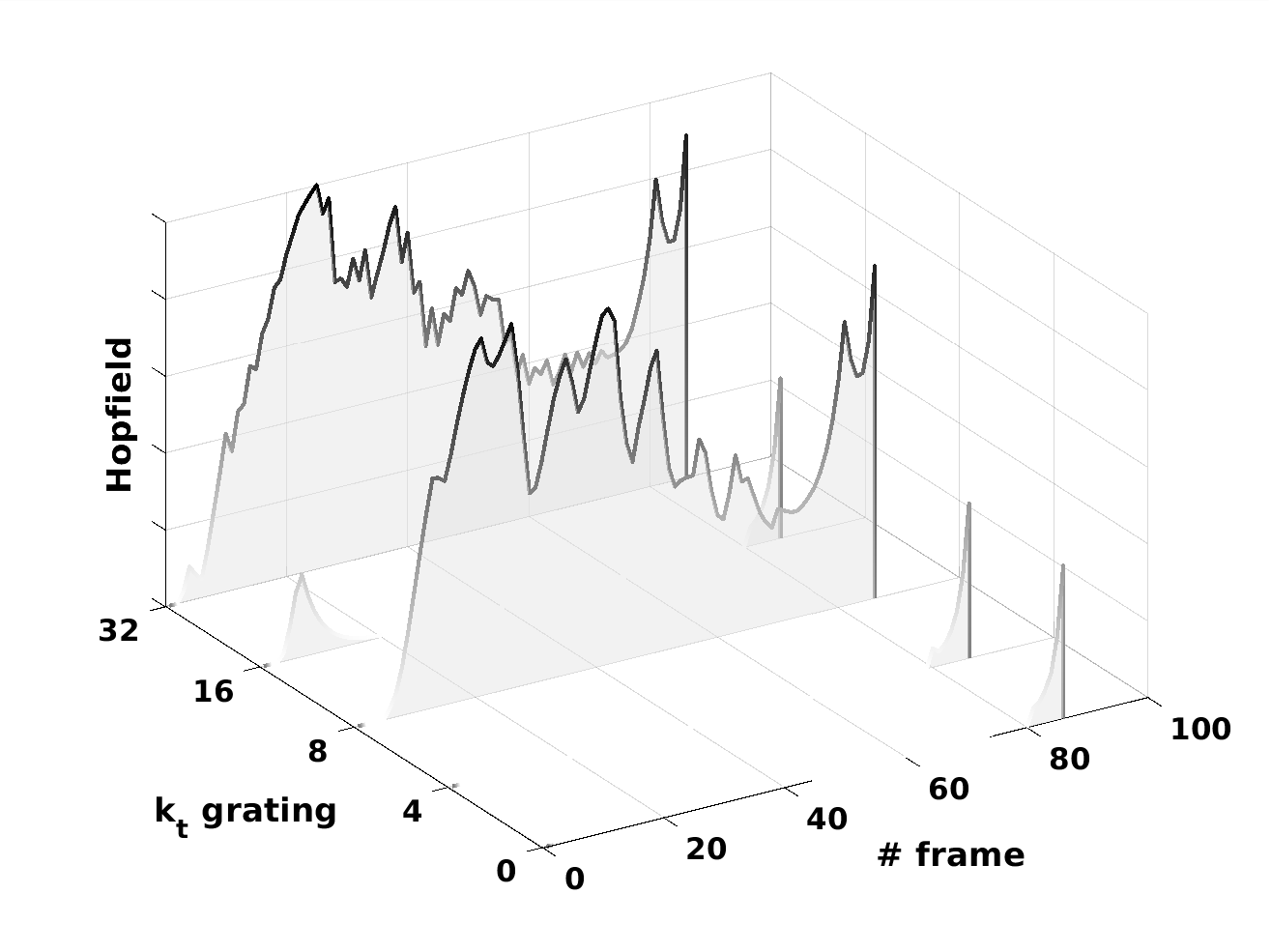}}
\caption{\label{FigDriftingSpeed}{\bf Drifting speed}.
The texture and approach parameter of the disk are identical with the previous figure.  The background consists of a horizontal sine wave grating with $k_s=16$ \cpix.  Its drifting speed was set to $k_t=0,\, 4,\, 8,\, 16\,\mathrm{and}\ 32$ Hz (axis labels).  The wave propagation vector points from the bottom to the top (arrow).
\textbf{(a)} Results for sum-of-temporal contrast (SOC).  Higher drifting speeds generate more temporal contrast.  For $16$ and $32$ Hz temporal contrast decreases at the end of the approach because the disk is occluding the grating.  A nearly identical plot would be obtained for $k_s=8$ \cpi (not shown).
\textbf{(b)}  Identical to Figure \ref{FigGratingOrientation}b, but this time with a horizontally oriented background grating.  The \hopfieldx-model shows resonance effects ($8,\, 32$ Hz vs. $4,\, 16$ Hz).  While responses to $k_t=0,\ 4\ \mathrm{and}\ 16$ Hz hardly contain usable information, the remaining two curves are noisy but show an increase in activity at the end of the approach.}
\end{figure}
For suitable combinations of $k_t$ and $k_s$, "resonance" effects due to temporal aliasing can be observed (Figure \ref{FigDriftingSpeed}b): The curves for $k_t=4,16$ Hz have less overall activity than those for $k_t=8,32$ Hz.  Thus, a nearly similar plot to Figure \ref{FigGratingOrientation}a would result when using $k_t=32$ Hz instead of $8$ Hz (not shown).\\
With a horizontally oriented grating as background, increasing the spatial frequency of the approaching disk generates less overall activity, especially in the initial approach phase (not shown). The resonance pattern observed with respect to $k_t$ is consistent with the one described above. With respect to the orientation of the approaching disk, activity in the initial approach phase tends to decrease close to the horizontal orientation. Again, for a vertically oriented background grating, retrieval activity is largely independent of disk spatial frequency and orientation.
Figure \ref{FigDriftingSpeed}a shows the sum-of-temporal-contrast (SOC, cf. Figure \ref{FigApproachCurves}) as a function of $k_t$. Since SOC is isotropic, there are no effects of disk or grating orientation. As SOC is a temporal high-pass filter, increasing $k_t$ increases SOC activity. Conversely, SOC is largely independent of $k_s$ (not shown).\\
With respect to the spatial frequency of the approaching disk's texture, SOC activity in the final approach phase increases with increasing spatial frequency. Thus, the final peak of the curves for $k_t=16$ and $32$ Hz in Figure \ref{FigDriftingSpeed}a could be recovered by increasing the disk's spatial frequency. No significant changes in the SOC curves occur when varying the disk orientation (not shown).
\begin{figure}[t!]
 \centering
 \subfloat[drifting foreground grating ($k_s$ varies)]{\includegraphics[width=0.5\linewidth]{\pics 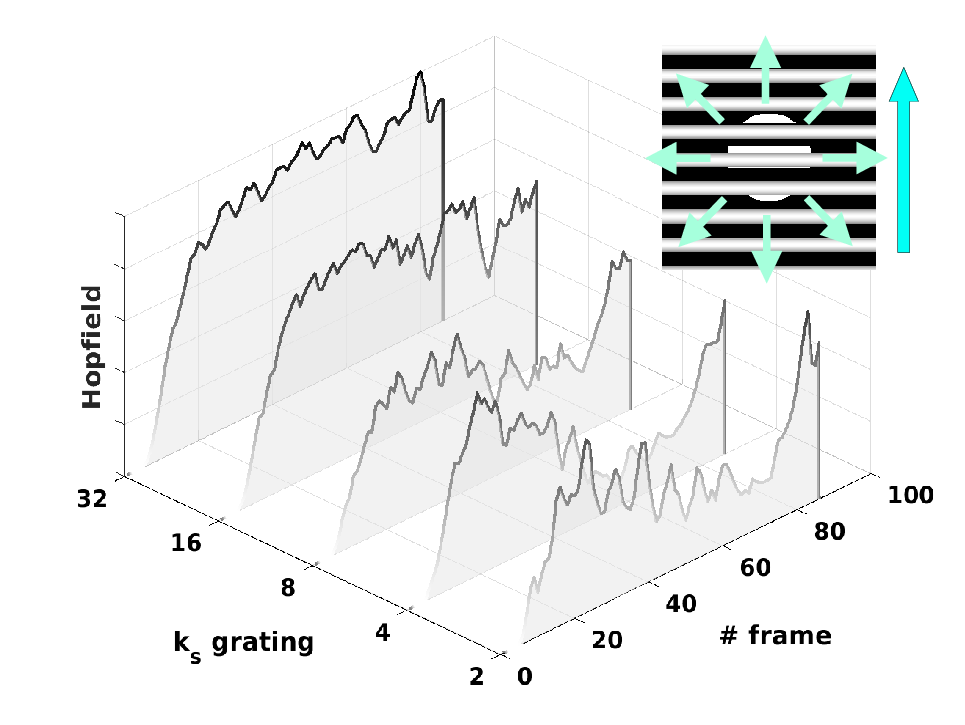}}~
 \subfloat[rotating grating ($k_s$ varies)]{\includegraphics[width=0.5\linewidth]{\pics 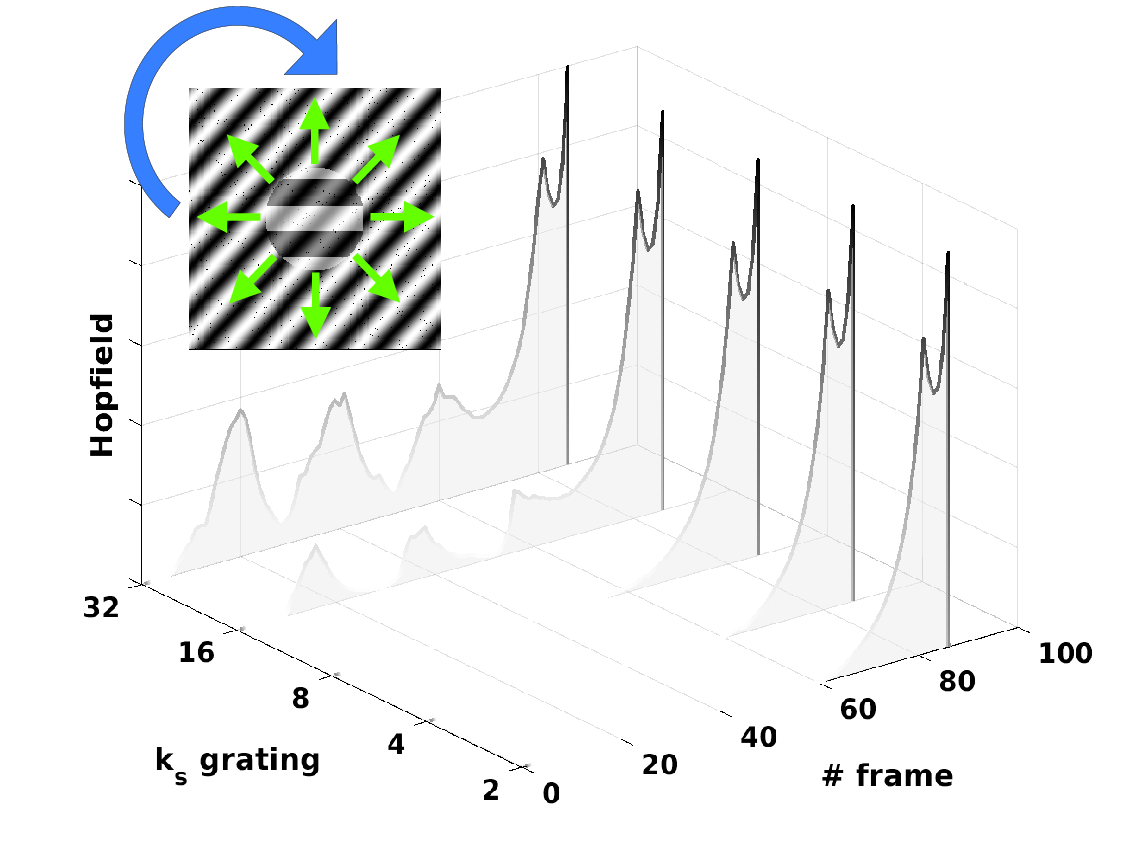}}
\caption{\label{FigGratingForegrndRotation}{\bf Foreground grating and rotating grating}.
\textbf{(a)} Same as Figure \ref{FigGratingOrientation}a, but with interchanged foreground and background: Here, the approaching disk is occluded by the bright bars of a drifting horizontal grating with $k_t=8$ Hz (see inset). Interference is increased compared to Figure \ref{FigGratingOrientation}a, leading to more overall retrieval activity, particularly in the initial approach phase.
\textbf{(b)} Analogous to Figure \ref{FigGratingOrientation}a, but for different spatial frequencies of a rotating background grating (without drift),  The background grating changes orientation with $8$ degrees per frame (after $180$ degrees it phase-reversed). With $86$ frames, the grating thus rotates nearly twice  during the approach. The approach was further complicated by \emph{(i)} randomly setting the pixels of each frame to zero with a probability of $0.01$, and \emph{(ii)} using a semi-transparent disk (i.e., alpha value $0.5$). The results for different rotating speeds (considered range $1$ to $32$ degrees per frame - not shown) are similar to those shown.} 
\end{figure}
When the drifting grating partially occludes the approaching disk (Figure \ref{FigGratingForegrndRotation}a), more spurious activity (interference) is generated than with the drifting grating in the background (Figure \ref{FigGratingOrientation}a). As before, interference only occurs when the orientations of the grating and the template pattern match (Figure \ref{FigPatternMemory}): a foreground grating with vertical orientation would produce similar results to those shown in Figure \ref{FigGratingOrientation}b for all spatial frequencies $k_s$ and drifting speeds $k_t$.\\
The spurious activity generated by a rotating background grating depends on both the rotation speed (in degrees per frame) and spatial frequency $k_s$. Generally, low rotation speeds (in the examined range from $1$ to $32$ degrees per frame) and higher spatial frequencies generate more interference. For low rotation speeds, interference can only occur if the grating orientation at some time $t$ is the same as that of the template pattern. Conversely, at high speeds, phase aliasing
may periodically generate spurious activity (see curve for $k_s=32$ \cpi in Figure \ref{FigGratingForegrndRotation}b). However, for all combinations of rotation speed (in the range of $1$ to $32$ degrees per frame) and spatial frequency of the grating, the activity increase in the final approach phase remains clearly distinguishable from the spurious activity generated before.\\
\begin{figure}[t!]
 \centering
 \subfloat[drifting horizontal grating]{\includegraphics[width=0.551\linewidth]{\pics 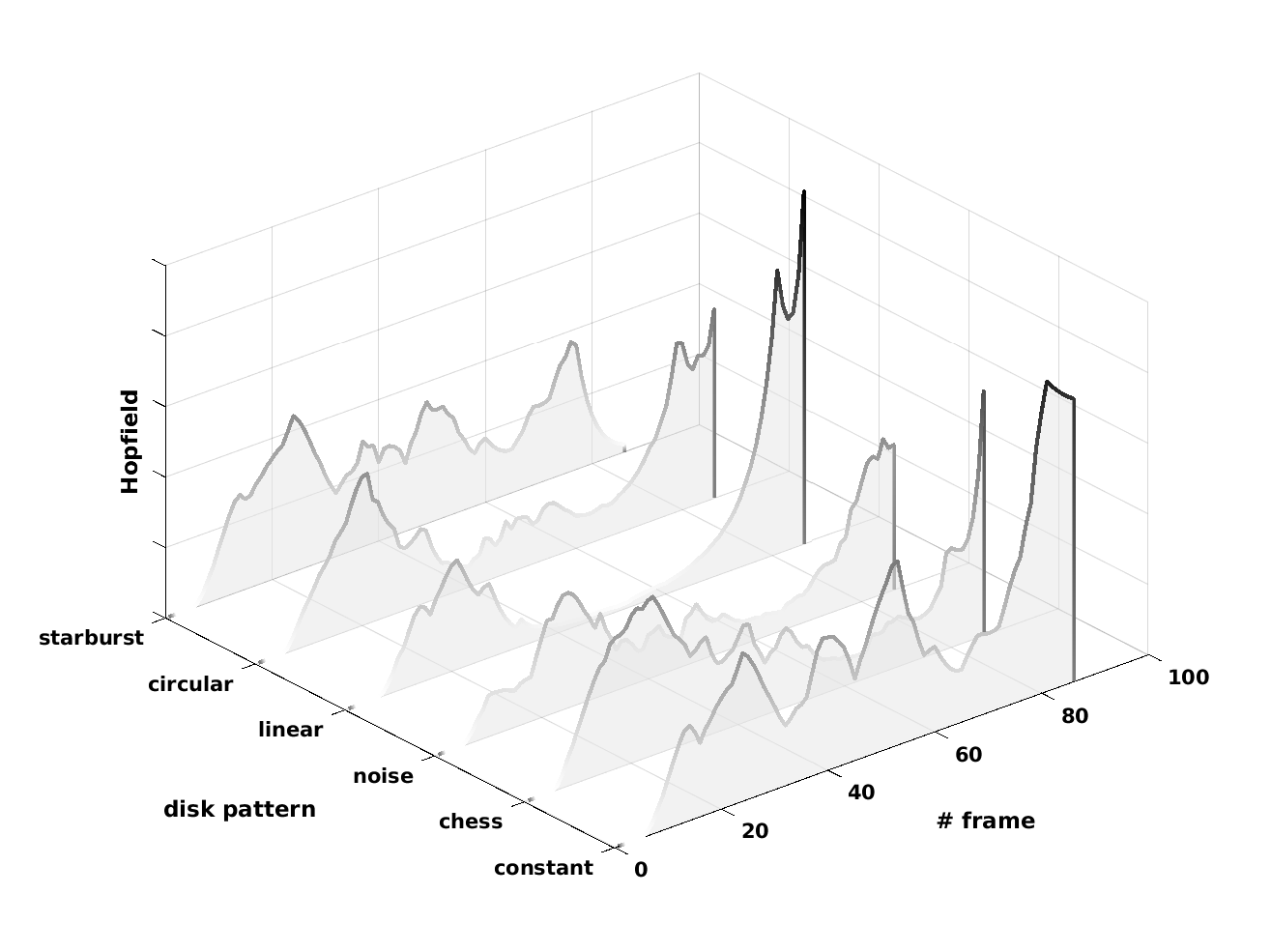}}
 \subfloat[object pattern]{\includegraphics[width=0.35\linewidth]{\pics 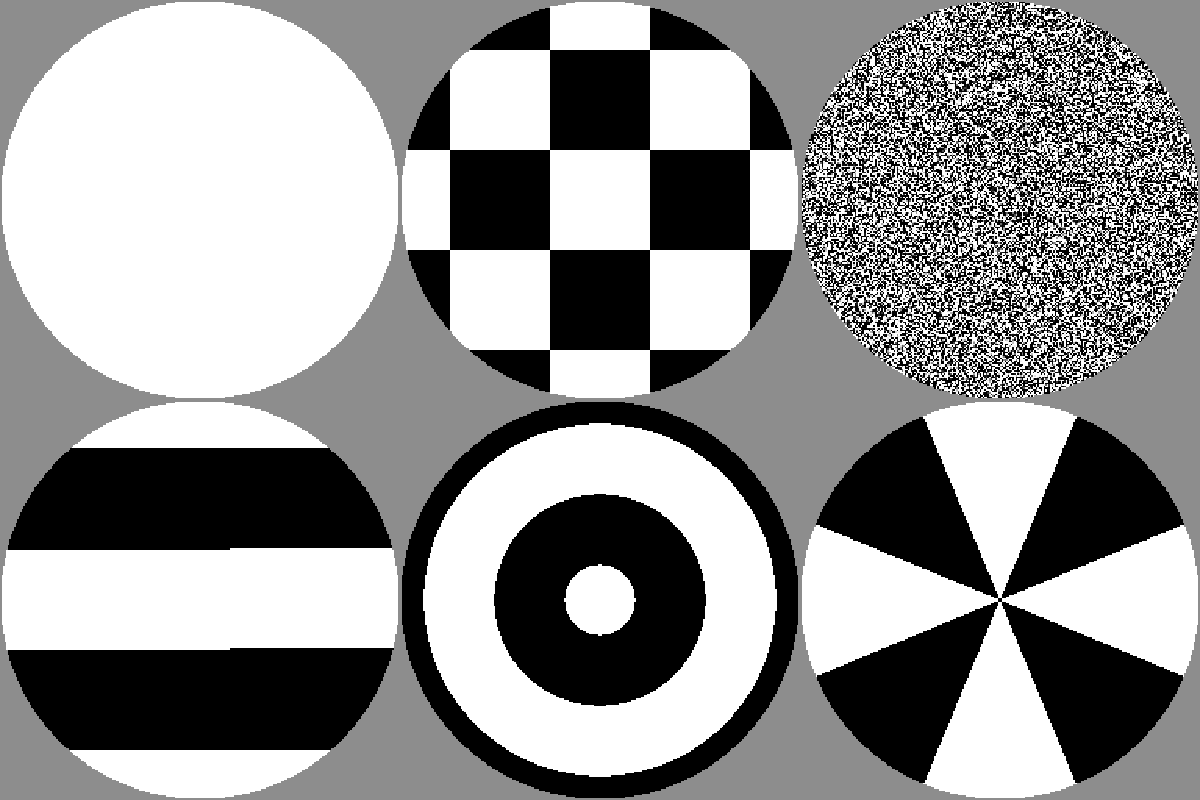}}~
\caption{\label{FigObjectPattern}{\bf Object pattern}.
\textbf{(a)} Responses of the \hopfieldx-model to the approaching disks as shown on the right (see axis labels). The background is a drifting horizontal grating ($k_s=8$ \cpi and $k_t=8$ Hz).  Approach parameters were the same as with Figure \ref{FigGratingOrientation}. 
\textbf{(b)} The approaching objects from left to right and top to bottom: uniform disk (axis label \emph{constant}); chessboard (\emph{chess}); \emph{noise} patch;  \emph{linear} grating (default); \emph{circular} grating; \emph{starburst}
grating}
\end{figure}
Can the \hopfieldx-model also signal approaching disks with a different pattern than the striped disk used as a template pattern (Figure \ref{FigPatternMemory})? To investigate this, we studied model responses with five additional object patterns (Figure \ref{FigObjectPattern}b) approaching over background gratings of various drifting velocities and orientations.  For a vertically oriented grating, the model responded to all object patterns except for the starburst grating.  The highest response amplitudes were obtained for the uniform disk and the linear (default) grating.  The chessboard pattern and the circular grating generated smaller amplitudes, while the smallest amplitude was obtained for the noise patch.  Responses did not significantly depend on the drifting speed or spatial frequency of the grating.  When using an approaching object with a square shape instead of a disk, similar responses were obtained for all six pattern.\\
When increasing the spatial frequency of the textured disks (except of the noise patch), then responses with more spurious activity are generated, and any response peak in the final approach phase gets smaller or disappears.  This implies that the detection of the corresponding approaching disk becomes virtually impossible.  As to SOC, the starburst grating and the uniform disk will generate practically no activity increase in the final approach phase (not shown).  All other types of approaching disks will generate a distinct peak in the final approach phase.  The peaks vary with spatial frequency, but nonetheless remain clearly visible (not shown).\\
With a diagonally oriented background grating, the gross response pattern as a function of drifting speed is analogous to that shown in Figure \ref{FigDriftingSpeed}b: For drifting speeds $k_t=0$ and $k_t=2$ Hz, very narrow response peaks are produced (i.e., very late response onset). For $k_t=4$ and $k_t=16$ Hz, responses do not contain spurious activity, but their onset occurs later than with the horizontal grating. For $k_t=8$ and $k_t=32$ Hz, response onset is similar to the horizontal grating, but responses are contaminated with small amounts of spurious activity.\\
Responses to the horizontally oriented grating resemble those of the diagonal grating, but with significantly more spurious activity generated at $k_t=8$ Hz (as shown in Figure \ref{FigObjectPattern}a) and $k_t=32$ Hz.\\
How do SOC responses behave with the mentioned object patterns and background grating configurations? SOC responses do not vary significantly with grating orientation and object shape (square vs. circular). Importantly, SOC shows a response peak in the final approach phase with the starburst pattern, where the latter and the uniform pattern have the smallest response amplitude compared to the rest. These peaks generated with the starburst and uniform object patterns are distinguishable up to $k_t=4$ Hz, and from $k_t \geq 8$ Hz, the activity generated by the background grating becomes larger, causing these peaks to disappear.  The response peak of the approaching linear (default) disk disappears for $k_t \geq 16$ Hz.  The SOC peaks of all other types of disks are eventually gone at $k_t = 32$ Hz.
%
%
%-------------------------------------------------------------------------------
\subsection{Model Shootout with Artificial Videos\label{SecArtificialShootout}}
%-------------------------------------------------------------------------------
%
%
The results of the sum of temporal contrast (SOC) and the \hopfieldx-model from the previous section can be attributed to their respective filtering characteristics: SOC is a temporal high-pass filter, and strong background movement can interfere with the detection of an approaching object. The \hopfieldx-model, meanwhile, uses spatial high-pass filtering and its performance depends critically on whether the background has a similar structure as the template pattern. If it does, spurious retrievals can be generated, making it difficult to detect the approaching object.\\
In this section, we compare the responses of the \hopfieldx-model and SOC with three other models that are based on temporal contrast ("SOC-based models" for short).  These are \yue (Section \ref{FuHuPe20} \cite{FuHuPe20}), \hex (Section \ref{CizekFaigl19} \cite{CizekFaigl19}), and \advanced  (Section \ref{MatsEliAngel04} \cite{MatsEliAngel04}).\\
As each of the models covers a different range of output values, their responses had to be normalized in order to be displayed in a single figure.  Specifically, the output of the \hopfieldx-model depends on the number of stored patterns $N$.  Because of Equation \ref{EqMultiplication}, the output can therefore vary between one and $N^2$.  The displayed curves were scaled by $z(t)/N^2$ times a factor that is specified in the figure legends where applicable.\\
For \yuex, we filtered the output spikes (cf. Section \ref{FuHuPe20}) which vary from zero to two.  For displaying, we accordingly divided the filtered output by two.\\
The rest of the curves (including SOC and the visual angle $\Theta$) were first divided by their respective maximum and subsequently multiplied by the maximum response between scaled \hopfield and scaled \yuex.\\
\begin{figure}[t!]
 \centering
 \subfloat[uniform disk]{\includegraphics[width=0.501\linewidth]{\pics 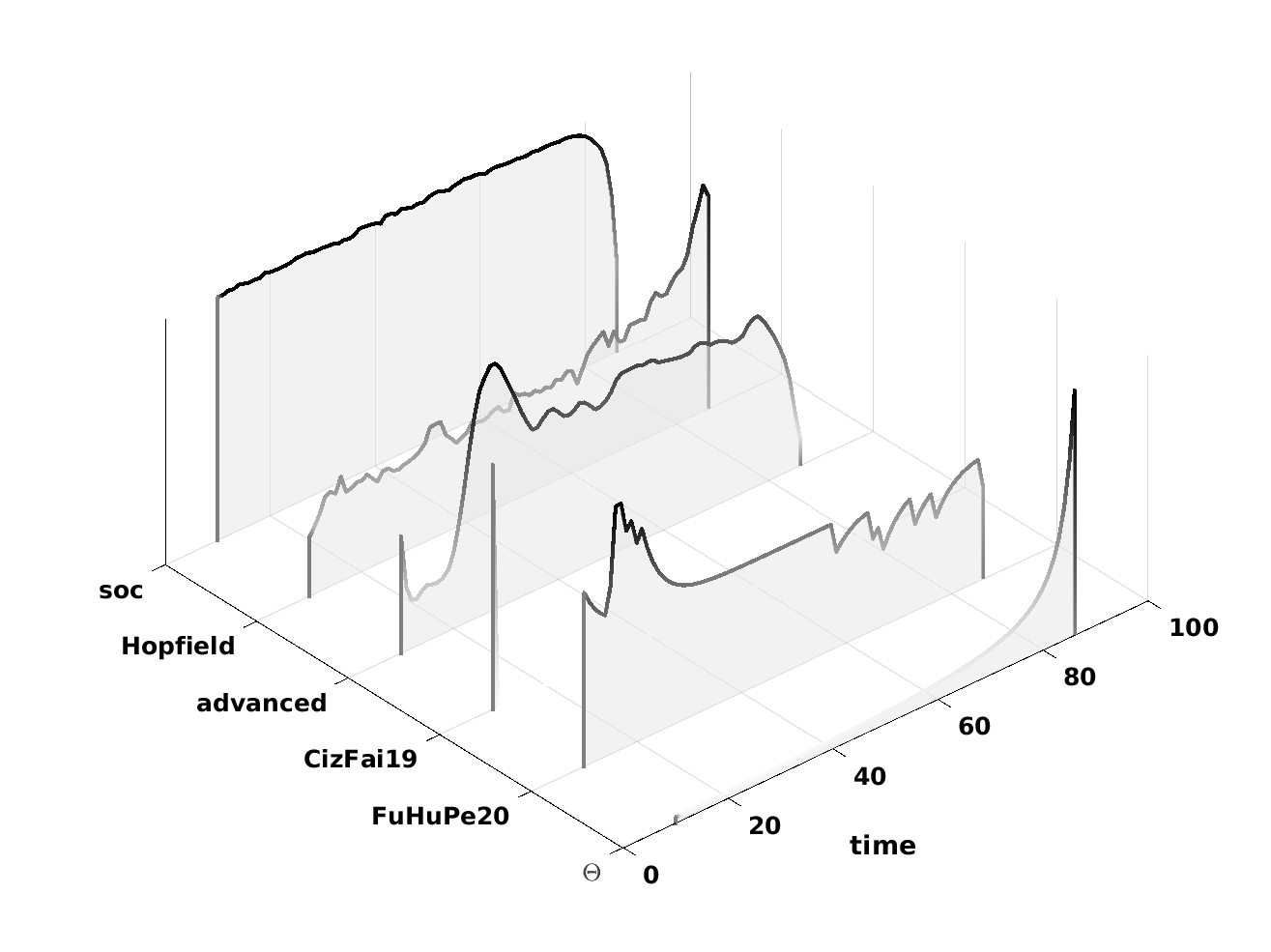}}
 \subfloat[noise patch]{\includegraphics[width=0.50\linewidth]{\pics 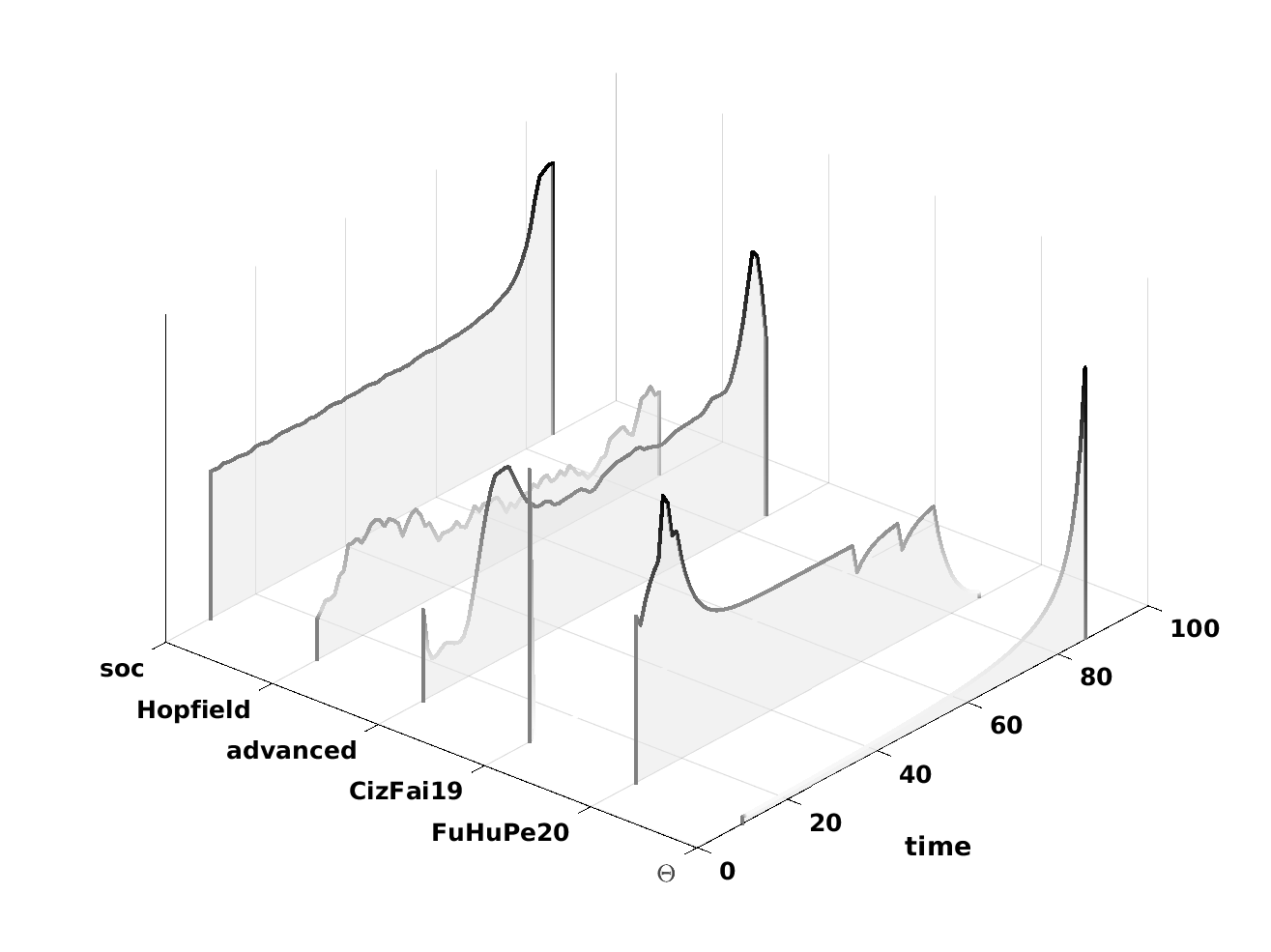}}~
\caption{\label{FigNoiseFloor}{\bf Dynamic noise floor}.  Response of the models described in section \ref{SecBenchmarkModels} along with the SOC and \hopfield to a disk that approaches against a noise floor (white noise image) that changes with each frame.  The physical parameters of the approaching disk are specified in Figure \ref{FigGratingOrientation}.  The normalized angular size $\Theta$ of the disk is also plotted (cf. Figure \ref{FigApproachCurves}).  For a better visibility, all curves start at $t=10$ and in this way initial transients are excluded.  The \hopfield responses were scaled by $2.5$.
\textbf{(a)} The disk had a constant luminance of $0.5$ (first disk in Figure \ref{FigObjectPattern}b), identical to the spatio-temporal average luminance of the background.  None of the models except of \hopfield show an increase in activity in the final approach phase.
\textbf{(b)} The disk consisted of white noise (third disk in Figure \ref{FigObjectPattern}b).  Unlike the background, the disk pattern remained the same throughout the approach.  Although SOC clearly encodes angular velocity, only one of the SOC-based models (\advancedx) reflects this activity increase in the final approach phase.}
\end{figure}
Figure \ref{FigNoiseFloor} juxtaposes model responses for two types of approaching objects (uniform disk and noise patch) against a background of dynamically varying white noise.  Luminance of the uniform disk was set to be identical to the spatio-temporal mean of the background, that is $0.5$.  This is likely a situation that would never be encountered in real-world applications, and it is even hard for humans on a calibrated computer monitor to perceive the approaching object before it nearly covers the entire frame.  The simulations suggest that it is difficult for the tested models as well.  With the uniform disk, only the \hopfield-model reveals an activity increase in the final approach phase.  SOC stays saturated until the uniform disk grows large enough such that it occludes the background.  This causes the activity decrease of SOC in the final approach phase.  As a consequence, none of the SOC based model signals the  approaching uniform disk.\\
A different situation is at hand with the approaching noise patch (third pattern of Figure \ref{FigObjectPattern}b), where SOC shows an activity increase with time.  Nevertheless, this increase is not mirrored in the responses of the SOC-based models except of \advancedx.  \hopfieldx, on the other hand, shows merely a moderate increase.\\
\def\foe{\mathit{foe}}	% focus of expansion
\begin{figure}[t!]
 \centering
 \subfloat[SOC]{\includegraphics[width=0.50\linewidth]{\pics 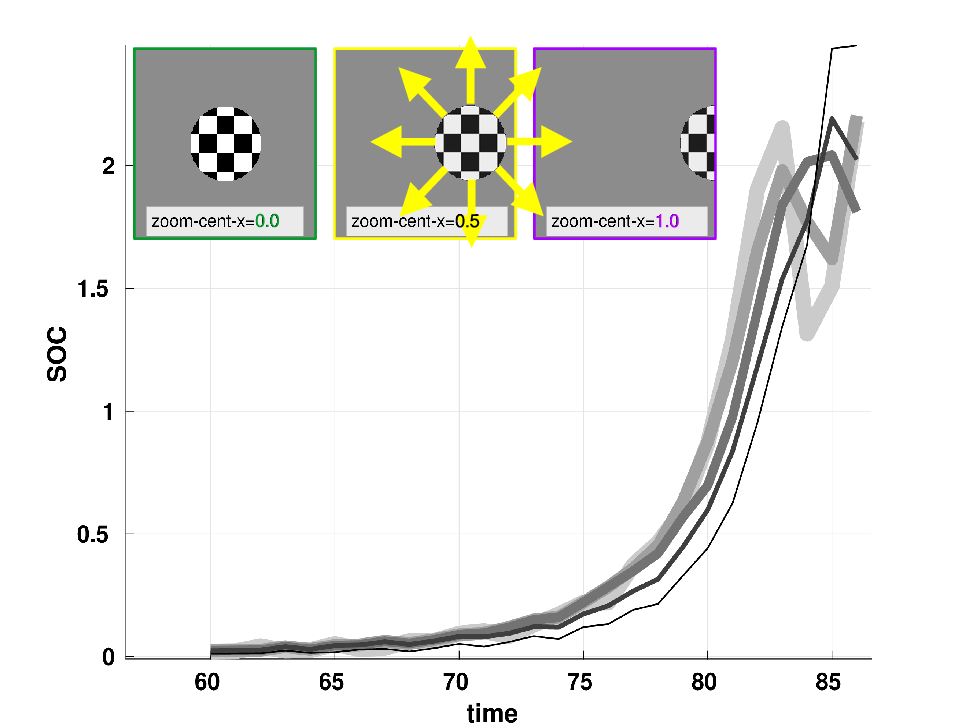}}
 \subfloat[Hopfield]{\includegraphics[width=0.50\linewidth]{\pics 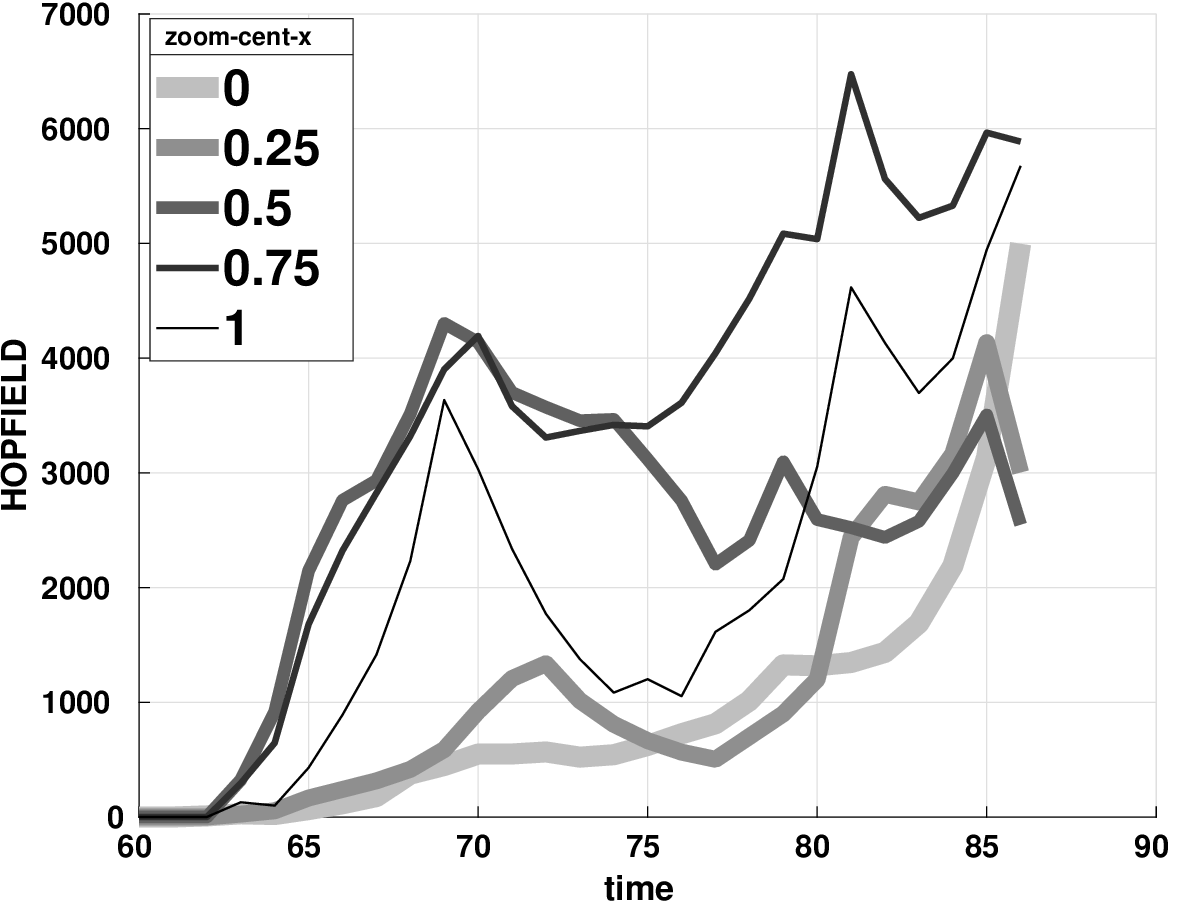}}~
\caption{\label{ZoomCenter}{\bf Focus of expansion}. The approaching object was a disk with a checkerboard pattern ($2$ cycles per object with full contrast, frame size $256 \times 256$ pixel) on a background with uniform luminance $0.5$ (medium gray).  The relative x-coordinate of the focus of expansion ("\emph{zoom-center-x}") varied from $x_\foe=0$ to $1$ in steps of $0.25$.  The relative values are transformed into absolute abscissa values by multiplying them with $256/2$ pixel.  The focus of expansion lies in the frame center for $x_\foe=0$  and at the right frame boundary for $x_\foe=1$ (see insets which shows sample frames at $t=73$ for $x_\foe=0,\ 0.5,\ \mathrm{and}\ 1.0$).
\textbf{(a)} The onset of SOC responses occurs later when shifting the focus of expansion to the right.  By and large identical behavior is observed for all SOC-based models and with different types of object patterns (e.g., an uniform disk).
\textbf{(b)} The responses of the \hopfield model at onset get steeper for increasing values of $x_\foe$.  The exact shape of the the response curves depends furthermore on the object pattern and background luminance.}
\end{figure}
How does the output of the various models depend on the location of the focus of expansion (FOE)? Figure \ref{ZoomCenter}a shows SOC which represent the input to the models \advancedx, \hexx, and \yue as a function of the horizontal position of the FOE.  The output of these SOC-based models follow their input, where each model's response occurs later with increasing displacement of the FOE from the center of the video frames.  This behavior is consistently observed irrespective of background luminance and object pattern.\\
In contrast, the output of the \hopfieldx-model is influenced by background luminance and object pattern. For example, the highest response amplitudes are generated for an uniform white or black background.  The closer the background luminance to medium gray, the smaller the \hopfieldx-responses.  Similarly, an approaching object with uniform luminance (e.g. a texture-less white disk) produces higher responses with an onset at an earlier time than a patterned object (e.g. a disk with checkerboard pattern).  When moving the FOE horizontally towards the frame boundary,  \hopfieldx-responses tend to increase stronger at earlier times compared to a centered FOE (Figure \ref{ZoomCenter}b).  Nevertheless, the time of the response onset does not change significantly. 
% .................................................................................
%
%-------------------------------------------------------------------------------
\subsection{Real-World Footage\label{SecRealWorldShootout}}
%-------------------------------------------------------------------------------
%
%
%
\begin{figure}[t!]
 \centering
 \subfloat[Mercedes]{\includegraphics[width=0.50\linewidth]{\pics 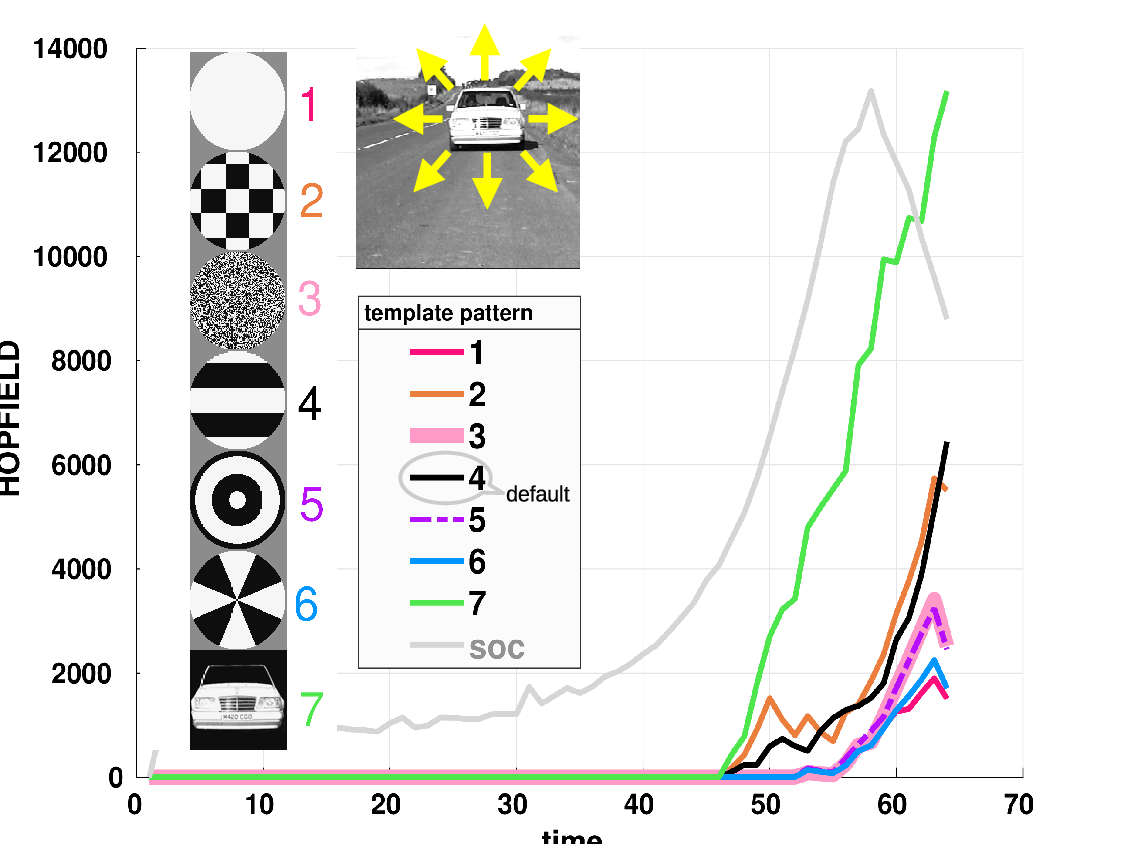}}
 \subfloat[Star Wars]{\includegraphics[width=0.50\linewidth]{\pics  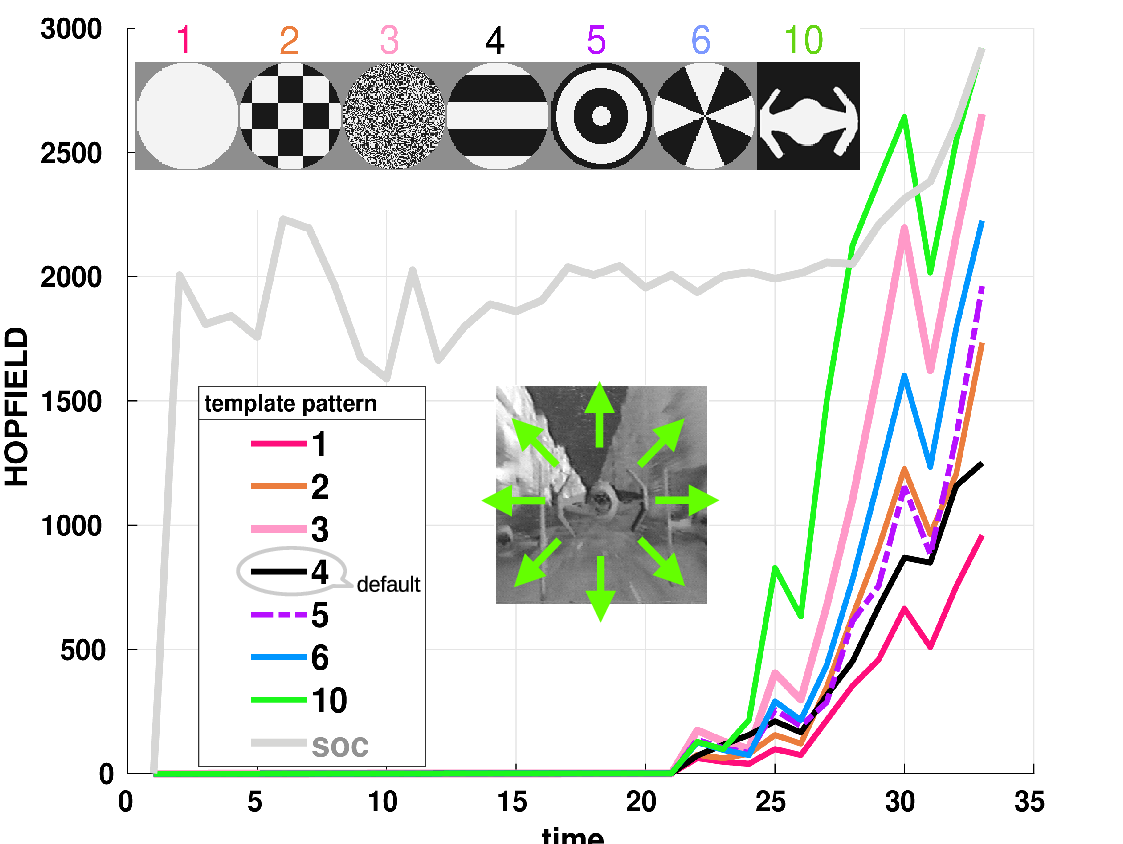}}~
\caption{\label{TemplatePatterns}{\bf Template patterns with real-world videos}.  The default template pattern corresponds to number four (cf. Figure \ref{FigPatternMemory}).  The plots shows \hopfield responses to real-world videos with alternative template patterns along with the sum of temporal contrast (SOC).  Both videos are courtesy of \emph{F.C. Rind} \cite{RindSimmons92a}.
\textbf{(a)} \emph{Mercedes Sequence}  ($64$ frames, frame size $285 \times 285$ pixel, inset shows frame number $32$).  The video shows an approaching car without background motion, yet with occasional camera shake of small amplitude.  Templates $2$, $4$ and $7$ have the earliest response onset. Template $7$ is the original car, and thus the optimal template pattern for this video.  It yields the highest response amplitude.
\textbf{(b)} \emph{Star Wars Sequence}  ($33$ frames, $285 \times 285$ pixels, inset: frame $17$; further frames are shown in Figure \ref{BallooncarStarwars}b).  Again, the highest amplitudes are obtained when the shape of the template matches the approaching object (template pattern $10$), followed by template $3$ (noise patch) and $6$ (starburst grating), respectively.  Notice that the SOC curve is rather noisy.}
\end{figure}
In this section model responses to four representative videos are compared with each other.  Firstly, Figure \ref{TemplatePatterns} shows responses of the \hopfieldx-model along with the sum of temporal contrast (SOC$=\sum_{ij}|\frame_{ij}(t)-\frame_{ij}(t- 1)|$) for  different template patterns.  The default template pattern is a horizontally-striped disk (Figure \ref{FigPatternMemory}), which gave the best overall results with a set of benchmark videos.  Therefore, the performance of the \hopfieldx-model may depend on the specific video under consideration.  Figure \ref{TemplatePatterns} shows that the optimal template pattern for each video is that which achieves the highest correlation with the approaching object.  This is a car for the \emph{Mercedes Sequence}, and the central space ship for the \emph{Star Wars Sequence}. The rest of the selected template pattern produced responses with lower amplitudes.  However, the next best (artificial) templates are not necessarily identical across different videos: While for \emph{Mercedes} these are the checkerboard disk and the default template, respectively, for \emph{Star Wars} we have the noise patch and the starburst grating, respectively.\\
\begin{figure}[t!]
 \centering
 \subfloat[Balloon Car]{\includegraphics[width=0.50\linewidth]{\pics 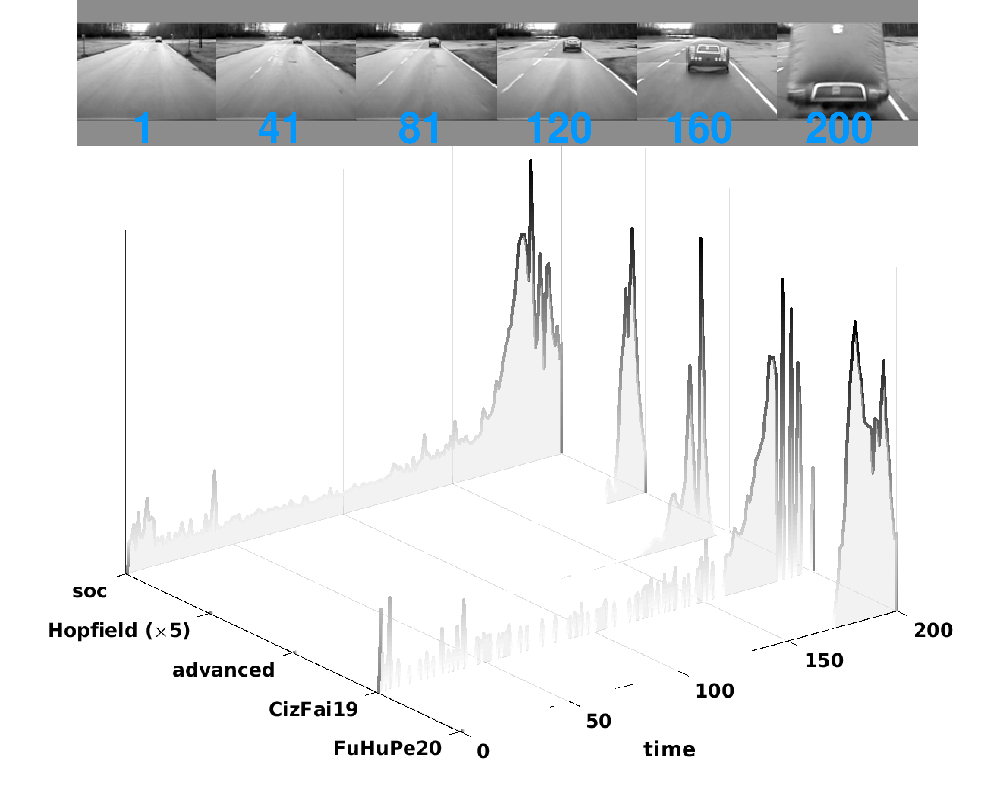}}
 \subfloat[Star Wars]{\includegraphics[width=0.50\linewidth]{\pics  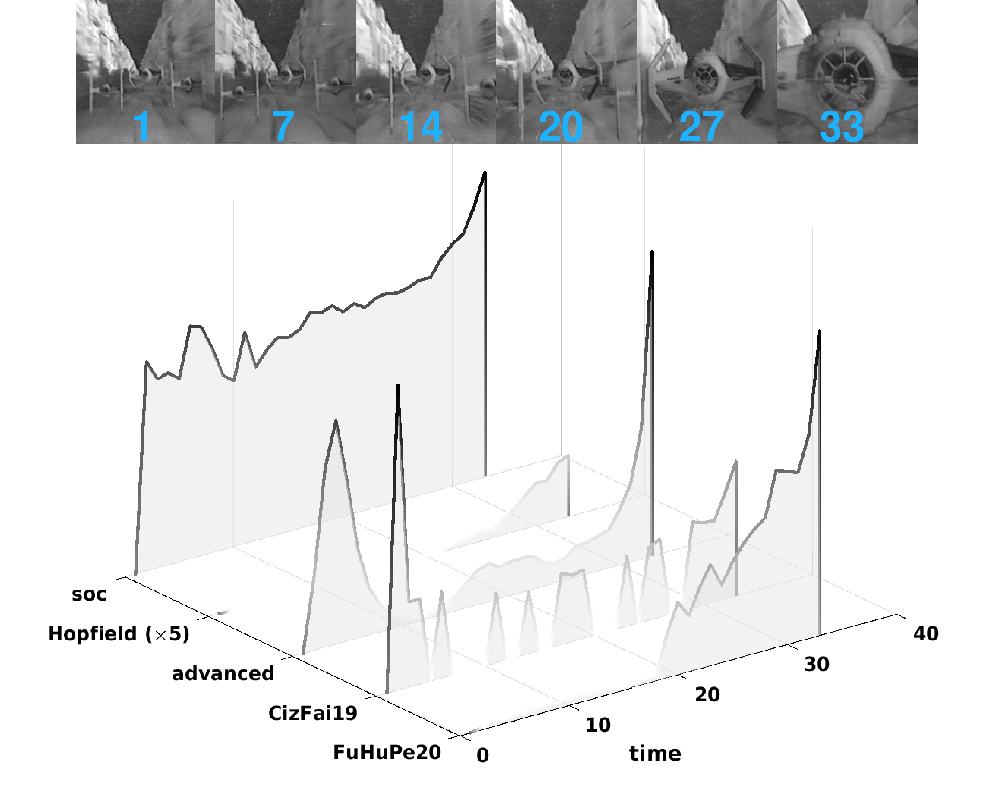}}~
\caption{\label{BallooncarStarwars}{\bf Benchmark Videos I}.  \hopfield responses were scaled by $2.5$.  With these videos, model responses should increase in the final approach phase.  Video snapshots at indicated frame numbers are shown at the top.
\textbf{(a)} \emph{Balloon Car Sequence} ($200$ frames, frame size $149 \times 149$ pixel, courtesy of \emph{Volvo Cars}).  The actual video size is $149 \times 98$ pixel, which was symmetrically embedded in a background with homogeneous luminance ($0.5=$ mid gray). The video shows a crash with an inflated mockup car.  The video contains smooth and moderate background motion. \textbf{(b)} \emph{Star Wars Sequence} ($33$ frames, $285 \times 285$ pixels, courtesy of \emph{F.C. Rind}).  The video has strong background movement in the opposite direction to the three approaching spaceships.  It has low luminance contrast along with occasionally random glitches.}
\end{figure}
Figure \ref{BallooncarStarwars}a juxtaposes responses of all models to the \emph{Balloon Car Sequence}.  The video has modest background movement and shows a crash with an inflated mockup car.  The background movement translates into non-zero SOC responses throughout the approach, and is well suppressed by the models \yuex, \advancedx, and \hopfieldx.  The response onset of \yue and \advanced coincide with the activity increase of SOC, where the latter two models respond earlier than \hopfield does. In summary, all considered models signal the approach to the stationary balloon car.\\
Figure \ref{BallooncarStarwars}b shows corresponding results for the \emph{Star Wars Sequence}, where three spaceships are approaching the observer. The observer looks into the opposite direction of his or her movement.  As a consequence, the background moves opposite to the approaching space ships as well.  This benchmark video is challenging due to strong background movement, low contrast, poor spatial resolution and occasional glitches.  The SOC activity reflects these properties, because activity is large, choppy, and increases during the whole approach.  As before, \yue and \hopfield are efficient in the suppression of background motion and start signaling the spaceship(s) in their final approach phase.  \advanced  shows a smooth and vigorous activity increase at the end of the approach, but still has non-zero activity before.  It also responds strongly to the first video frames until adaptation to background motion is completed.  Finally, \hex has a rather jaggy output, and the activity increase at the end is probably not sufficiently pronounced for a consistent detection of the imminent collision.\\
\begin{figure}[h]
 \centering
 \subfloat[Train]{\includegraphics[width=0.50\linewidth]{\pics 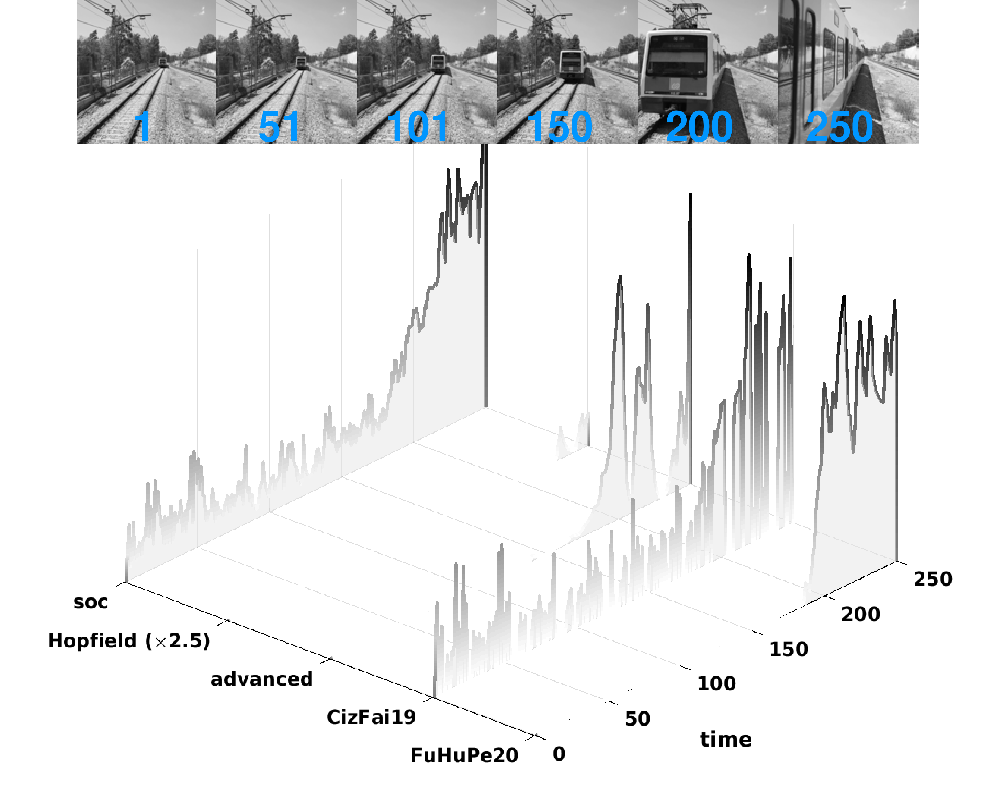}}
 \subfloat[Highyway]{\includegraphics[width=0.50\linewidth]{\pics  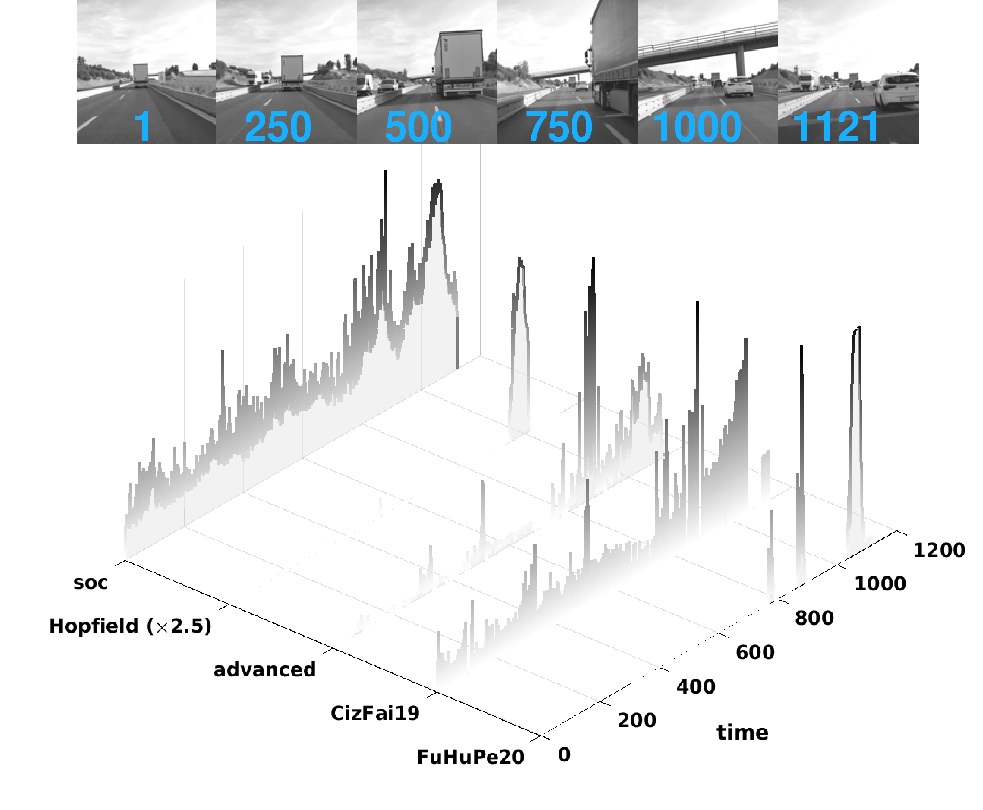}}~
\caption{\label{TrainHighway}{\bf Benchmark Videos II}.  \hopfield responses were scaled by $5$.  With both videos, no model responses should be observed.
\textbf{(a)} \emph{Train Sequence} ($250$ frames, frame size $256 \times 256$ pixels).  The video shows a passing by train and involves occasional camera shake.  Since the train does not collide with the observer, the models should ideally ignore it.
\textbf{(b)} \emph{Highway Driving}  ($1121$ frames, $256 \times 256$ pixels).  The video shows typical highway driving and contains moderate camera shake.  It involves overtaking a truck first (around frame $500$), and then driving under a bridge (around frame $1000$).  Notice that many commercial collision detection systems would signal false alerts to approaching bridges or tunnels.}
\end{figure}
Figure \ref{TrainHighway} shows model responses to two videos without an actual or imminent collisions.  Therefore, all models should ideally stay silent.  The first video shows a train that seems to be on a collision course, but ultimately drives past the viewer.  Because the camera was handheld, there is also a certain amount of camera shake which accounts for the non-zero SOC responses throughout.  Camera shake translates to sudden wide/large-field movement which should be suppressed.  In the locust visual system, LGMD responses to wide-field motion are suppressed by feedforward inhibition \cite{RindSimmons92a,RindBramwell96}.  From all models, \hopfield shows the smallest activity (note that \hopfieldx-responses were scaled by factor $5$).  The rest of the models have a clear increase of activity in the final phase of the approach, where wide-field movement is suppressed in the output of \yue and \advancedx, but not in \hexx.\\
Figure \ref{TrainHighway}b shows two typical situations that may occur when driving on a highway: first we overtake a truck and then we drive under a bridge.  None of these events implies an imminent collision, therefore all model output should remain zero.  The video involves complex background movement (e.g. oncoming vehicles, road marking, guard railings), as evidenced by the noisy SOC activity.  SOC also reflects the overtaking maneuver from frame $500$ to $900$.  The first activity peak (frame $500$) is when the rear of the truck moves out of sight.  The second peak (frame $900$) comes from the front of the truck as it moves out of the field of view.   Finally, the last peak is produced by the bridge.  The \hopfieldx-model in particular does not respond to these events, but shows a spurious peak around frame $1000$, just when the overtaking maneuver has been completed (recall that \hopfieldx-responses were scaled by $5$).  It stays silent during the rest of the drive. The \yuex-model shows narrow response peaks to all events: start and finish of overtaking and the bridge.  \advanced shows a strong response to the end of the overtaking maneuver, and less to the bridge. There appears another narrow peak with low amplitude around frame $300$, caused by an oncoming truck on the opposite lane.  Finally. \hex responds to all events in an undifferentiated way.
%
%
%--------------------------------------------------------------
\section{Discussion and Conclusions\label{ Conclusions}}
%--------------------------------------------------------------
%
%
\textbf{Summary.} In this paper, I proposed a radically different algorithm for signaling object approaches which is based on modern Hopfield networks.  It includes separated information processing along parallel ON- and OFF-channels.  The critical mechanism that enables the use of Hopfield networks in the context of collision detection is a dynamic pattern memory.  That is, at each time step,  the memory is updated with a delayed video frame.  Since Hopfield networks will always retrieve the pattern that best correlates with its initial state, the model would perform rather erratically without the memory update.\\
The proposed model is quite different from the bulk of published approaches, which extract temporal contrast from their input.  SOC-based models are often biologically inspired, and model, for example, the neuronal circuitry of the LGMD neuron of the locust.  To assess the performance of the \hopfieldx-model, its principal characteristics were studied with artificial video sequences.  Subsequently, the output of three representative SOC-based models where compared with \hopfield for a couple of benchmark videos.\\
All of the considered SOC-based models are highpass filters in time, and their output is related to the angular velocity (cf. Figure \ref{FigApproachCurves}).  Conversely, since in the memory of the \hopfieldx-model a template pattern is stored with different sizes, its output is related to angular size.  Furthermore, \hopfield involves spatial highpass filtering for processing the video frames and for its memory.\\
The results of the models can often be attributed to their different filtering characteristics (spatial vs. temporal highpass).  For instance, SOC-based models will eventually fail to track an approaching object against a background grating with a high drifting frequency.  \hopfieldx, on the other hand, will have a poor performance when the background grating matches the orientation of the template pattern (cf. Figure \ref{FigPatternMemory}).  In that sense, both model types are complementary and could be used in parallel in order to reduce the number of false collision alerts or missed approaches.\\ 
It is likely that the \hopfieldx-model could be made more robust by increasing the number of pattern memories for storing a variety of template pattern.  This was not subject of the present work, since it has to be studied carefully how to combine the retrieval results across multiple pattern memories.\\
\textbf{Dashcam footage and cropped video frames.} In reference \cite{Hartbauer17}, dashcam videos with different traffic situations were shown to locusts and responses of the descending contralateral movement detector (DCMD) neuron were recorded.  The DCMD replicates LGMD firing up to a frequency of $400$ Hz \cite{Rind84,JudgeRind97}.  It was found that the DCMD could distinguish traffic situations where collisions occur from normal driving.  After challenging the \hopfieldx-model with dashcam footage extracted from Youtube crash compilations, I found that the \hopfieldx-model did not generate any response to $25$ videos showing all types of collisions (e.g., frontal and lateral) and near misses.  The dashcam videos usually have a wide field of view and have much distortion which is typical of non-corrected lenses in short focal lengths.  Furthermore, due to the aspect ratio, the footage typically occupied only one third of the frame when embedded in a square frame format.  The \emph{Balloon Car Sequence} of Figure \ref{BallooncarStarwars}a has similar properties, but a narrower field of view with few distortion.  Furthermore, an uncropped version of the \emph{Balloon Car Sequence} did not change the responses of \hopfieldx, nor did the \hopfieldx-model respond in a different way to a cropped version of the \emph{Train Sequence} (Figure \ref{TrainHighway}) with a wider field of view.  This suggests that the field of view and the degree of distortion of the video is critical for a proper functioning of \hopfieldx.\\
\textbf{Usage of stored patterns.} The amplitude of \hopfieldx-responses reflects the retrieved pattern.  Why does the model only use a limited range of the stored pattern vectors, but not continuously respond to the angular size of an approaching object?  This is because of the dynamic memory update by a five time steps delayed video frame (Equation \ref{EqOnOffMemory}).  For a ``vanilla'' object approach (uniform white disk approaching against a uniform black background, see Figure \ref{FigApproachCurves}), the relationship between the delay and the response onset is such that the longer the delay, the earlier the response onset to the approaching disk.  However, the advance of the response onset is only moderate, even for long delays (e.g. $5$ vs. $20$ time steps).  This relationship does not readily generalize to real-world videos, where with different numbers of delay time steps the response curves of \hopfield may become noisier, response amplitudes may be altered, and finally the response onset may be advanced or even delayed.  Furthermore, when an object that will eventually collide with the observer is still far, then it may not be exactly in the image center (e.g. due to camera shake).  As consequence the tiny versions of the template pattern will not be retrieved.  Therefore, for real-world videos one can expect lower hit-rates at bigger distances.  Notice that distant objects will also generate few temporal contrast, leading to a comparable ``problem'' for SOC-based models as well.\\
\textbf{Biological plausibility.} In humans, (auto-) associative processes are thought to play an important role for perception, prediction and behavior \cite{BarAminoffMasonFenske07}.  For example,  object recognition in the brain makes use of such content-addressable memories in order to organize bottom-up sensory input \cite{Bar04}. This makes the whole process quick and reliable, even when the sensory input is incomplete (e.g. occluded objects), ambiguous (e.g. few visual cues \cite{TeufelDakinFletcher18}) or has a poor resolution (e.g., face recognition from a large distance \cite{CoxMeyersSinha2004}).  Therefore, at least for the human brain, \hopfield may represent a plausible model of how angular size is perceived.  Note that when the extent of an object has been learned previously, then its exact absolute distance could be computed from measuring the angular size of its image projected on the retina.  The pattern memory (Equation \ref{EqOnOffMemory}) may be implemented biologically with a different structure than a matrix.\\
To simplify notation, I refer to retinotopically arranged units such as photoreceptors and postsynaptic neurons (e.g. motion detectors \cite{HasRei56}) as \emph{image} for short.  The pattern memory could be implemented with a \emph{dendritic processing scheme} as follows.  Synapses at the distal part of the dendrite would receive input from the central part of the image.  The connectivity pattern would replicate the shape of the template pattern (Figure \ref{FigPatternMemory}).  Since the template patterns are sparse (because of spatial highpass filtering), connections would only be required along their contours.  Synapses at the proximal part of the dendrite would receive input from the peripheral regions of the image.  With this scheme, an approaching object would activate each time more proximal synapses as it moves closer to the observer.  At the start of the approach, a postsynaptic potential (PSP) is generated at the distal site.  This PSP package travels down the dendrite while the object is approaching further and keeps on with generating more PSPs along the dendrite.  If the location of the synapses and the length of the dendrite is matched with the preferred speed and size of the approaching object, then previously generated PSPs (that travel downwards) coincide with currently generated ones. This would cause a continuous increase of the PSP amplitude during an approach.  A suitable chosen threshold on the final PSP amplitude would therefore be tantamount to a threshold in angular size.\\
In fact, many animals and insects appear to trigger escape or avoidance reactions by a threshold in angular size \cite{TomRin23}.  The Fiddler crab seems to be an exception as it relies instead on a threshold in angular velocity \cite{DoBaPa22}. 
Specifically, it has been hypothesized that locusts optimized their escape reactions to predators of diameters around $50$ to $90$mm \cite{RindSanter04}, and avoidance reactions in flight are triggered around $10$ degrees of visual angle \cite{RobJoh93b}.  Any such size preference must be encoded somewhere in the LGMD circuit, and the just outlined dendritic processing scheme could master that.\\
Functions like $\tau\equiv \Theta/\dot{\Theta}$ for explaining distance perception \cite{Gibson1950} or $\eta\equiv\dot{\Theta}\exp(-\mathit{const}.\cdot \Theta)$ for fitting LGMD-responses \cite{HatsopoulosEtAl95} postulate the availability of angular size $\Theta$ apart from angular velocity $\dot{\Theta}$ (i.e., SOC).  So do theoretical accounts which address the biophysical implementations of $\tau$ and $\eta$, respectively \cite{GabKraKocLau02,NIPS2011_0348,MatsJoan2012,Mats2015}.  Although non-retinotopic feedforward inhibition received by the LGMD seems to be related to $\Theta$ \cite{Palka67,RowellEtAl77}, no precise statement has been made as to its computation.\\
On the other hand, the majority of models for collision detection and avoidance start with computing the difference between successive image frames.  As mentioned, summing the activity across a difference image is proportional to angular velocity in the absence of both background movement and weird lighting conditions.  Mathematically, in order to recover angular size, one has to integrate successive measurements of angular velocity.  Integration on-the-fly could be carried out by a dendritic processing scheme similar to the one sketched above.  Another possibility to estimate angular size (given rate of expansion) is by suppressing all activity enclosed by the outer contour of an object's projected image \cite{MatsEliAngel04}.  The idea therefore is to just keep the activity corresponding to the outer contour.  In the case of a sphere, the outer contour is the circumference of a circle.  Then, the mean activity across the image would be proportional to angular size.\\
\def\co2{$\mathrm{CO}_2$}
\textbf{Learning to avoid obstacles} 
Reinforcement learning (RL) lends itself to detect or avoid collisions: a reward could be issued upon successfully avoiding an obstacle.  Otherwise a penalty is imposed.  Usually, RL entails the stochastic exploration of a set of rules (policies) for achieving some goal or task.  Model parameters are adjusted according to the reinforcement signal.  Abstractly speaking, however, RL just optimizes model parameters.  In this broader sense, RL could appear in different scenarios.\\
For example, in this paper $2370816$ combinations of eight model parameters were systematically explored.  Subsequently, the parameter values were selected which achieved the best possible performance over a set of eight benchmark videos. The selection of model parameters according to some evaluation score (or fitness function) is also at the heart of genetic algorithms (GA; \cite{Mitchell96}).  Genetic algorithms converge faster to a good solution compared to brute-force-parameter-parsing.  However, there is no guarantee that a GA finds the best solution.\\
A GA with a population size $40$ was used in \cite{YueEtAl06} to optimize six model parameters of a simple LGMD-model .  The optimization criterion was to reduce the number of false alerts and false misses across a set of benchmark videos.\\
In robotics, obstacle avoidance is often combined with path selection and navigation methods.  Specifically, deep reinforcement learning (DRL) allows to train a single neural network that uses video images as input and motion signals as output (end-to-end approaches).  Hierarchized architectures were proposed as well, where sensory processing and navigation is separated.  Since it is impossible to give a in-depth review on this rapidly evolving topic, I limit myself to highlight a couple of typical examples.\\
The approach proposed in \cite{GotShi17} combined a chaotic neuronal network (the \emph{actor}) with a regular neuronal network (the \emph{critic}).  The critic evaluates the actor's performance and computes the reward signal for RL.  Both the actor and the critic received a total of $146$ sensory signals which informed about locations of the obstacle and the target, respectively, and the distance to the walls.  The output of the actor are motor commands for the robot (left / right).  Although the environment and problem configuration was rather simple, the interesting aspect of the proposal lies in the generation of the stochastic movements of the agent.  With usual RL techniques, external noise has to be applied to generate the random behavior which is rewarded or penalized.  In  \cite{GotShi17}, however, the external noise was replaced by the internal dynamics of the chaotic neuronal network.  Thus, during the learning process, attractors may form according to the agent's goal.  The network therefore can be tuned to be more goal-directed or more exploratory.\\
The approach of \cite{HeAoWh20} used an LGMD model from \cite{YueRind06}.  Rather than plain luminance. normalized image moments \cite{Hu1962} were fed into the LGMD model.  Image moments are less sensitive to  camera noise and intensity variations, respectively.  The output (one dimensional) of the LGMD model was fed into a deep neuronal network (DNN) along with the relative position to the target to which a micro unmanned aerial vehicle (UAV) should move.  The DNN was trained with DRL where an explicit reward function was used. It outputs navigation commands for the micro UAV.  The trained network navigates the UAV through complex environments and thus shows that the one-dimensional LGMD signal computed from monocular camera input is sufficient for successfully avoiding collisions.  This is remarkable in the sense that no explicit depth information seems necessary.\\
The latter proposal stands in contrast to \cite{WeScLe21}, which relied on estimating depth information.  To this end, a generative adversarial network (GAN) was trained to predict depth maps from monocular camera images in "simple, maze-like environments".  An end-to-end approach was taken (where several models were compared to each other), where the input were the camera images along with its predicted depth map.  For training a laser range finder was employed in order to determine the reward signal that was computed with an explicit reward function.  Despite of being a generative model, the depth predictions of the GAN will likely fail in complex and unconstrained environments.\\
In  \cite{ZhLiKi22}, a shallow network was trained to replicate the receptive field (RF) structure and response properties of \textit{Drosophila's} LPLC2 neurons  \cite{KlapoetkeEtAl17}.  Receptive fields (RFs) were represented as $12 \times 12$ kernels ($=$ network output), where two RF models were compared with each other: \emph{linear receptive field} (LRI) units and \emph{rectified inhibition} (RI) units, respectively.  The input to the network was optical flow magnitude in four orthogonal directions.  Training data consisted of four types of artificial motion pattern (``loom-­and-­hit'', ``loom-­and-­miss'', ``retreat'', ``rotation'') and were labeled with their respective collision probability (one for ``loom-­and-­hit'', zero for the rest).  A total of $4000$ trajectories were generated for training.  The filter kernels were evenly distributed across visual space. Different networks were trained with a different number $M$ of kernels ($M=1,2,4,\ldots,256$). Individual filter responses were pooled to evaluate the overall performance for signaling collisions.   By assuming circular and mirror-symmetric kernels, the number of trainable parameters was reduced to $56$ (LRF) and $112$ (RI), respectively.  Three further kernel types were created by rotating the trained kernel by $90, 180$ and $270$ degree.  The kernels were arranged accordingly across visual space.  The resulting kernels matched their biological counterparts in that their pooled response is sensitive to radially outward moving edges such as being generated during an object approach, but is inhibited by retreating objects.  Apart from this outward solution, a trivial solution and an inward solution emerged as a result from several training sessions with different network initializations.  In line with biological LPLC2 neurons \cite{AcPoAl19} the pooled responses encode angular size, although the input corresponds to optical flow signals (i.e., directional angular velocity).\\
In summary, there are two conceivable scenarios for the use of deep reinforcement learning (DRL) to avoid collisions.  First, one can train a DRL-architecture with the output of any proposed collision avoidance model (CAM for short) to predict from their output whether a collision is about to occur or not (e.g. similar to \cite{HeAoWh20}).  Alternatively, the DRL-architecture could be trained to output the probability of an imminent collision.  The second scenario is an end-to-end approach, which uses video frames as input (e.g. analogous to \cite{WeScLe21}).  The advantage of the first approach is that it is less computationally demanding than an end-to-end approach.  It can be expected that a vision-based end-to-end approach that reliably works in unconstrained environments requires a huge amount of (labeled) training data.\\
In general, although the performance of deep learning (DL)-architectures is often spectacular, the energy demand for training should not be undervalued.  For example, reference \cite{StGaMc19} estimated that the development of an DL-architecture up to publication standards typically requires the training of $4789$ candidate models across six months, what amounts to more than $35$ kilotons of \co2 emissions. Even worse, training large models such as a transformers will generate about $284$ kilotons of \co2.  This should be compared to the relatively modest computational demand of most published CAMs for development and optimization.  Moreover, by adjusting frame rate and/or frame size, all of the considered CAM models can run on current hardware in real-time.   A further concern about current DL-architectures relates to reliability and predictability. It is well established that DL-applications reflect any bias inherent in the data which were used for training.  With respect to generalization performance, the complexity (i.e. the very number of free model parameters) disallows any systematic analysis based on its final configuration (i.e., after training).  Apart from providing only limited insights into the information processing chain of a trained network, rather unexpected failures were reported: for example, DL-architectures trained for traffic sign recognition (or the recognition of other objects) can easily be led astray by just applying small modifications to the original traffic signs (e.g. with a sticker or a marker) \cite{SitawarinEtAl18,EykholtEtAl2018,OnePixelAttack2019}. Suchlike modifications would not impair the performance of CAM models.

\begin{acknowledgements}
This study was financially supported by grant PGC2018-099506-B-I00 and PID2022-142599NB-I00 from the Spanish Government.
\end{acknowledgements}

 {\small
%\bibliography{./refs}{}

\begin{thebibliography}{10}

\bibitem{AcPoAl19}
Jan~M. Ache, Jason Polsky, Shada Alghailani, Ruchi Parekh, Patrick Breads,
  Martin~Y. Peek, Davi~D. Bock, Catherine~R. {von Reyn}, and Gwyneth~M. Card,
  \emph{Neural basis for looming size and velocity encoding in the drosophila
  giant fiber escape pathway}, Current Biology \textbf{29} (2019), no.~6,
  1073--1081.e4.

\bibitem{Bar04}
M.~Bar, \emph{Visual objects in context}, Nature Reviews Neuroscience
  \textbf{5} (2004), 617--629.

\bibitem{BarAminoffMasonFenske07}
M.~Bar, E.~Aminoff, M.~Mason, and M.~Fenske, \emph{The units of thought},
  Hippocampus \textbf{17} (2007), no.~5, 420–428.

\bibitem{BlaRinVer99}
M.~Blanchard, F.C. Rind, and F.M.J. Verschure, \emph{Collision avoidance using
  a model of locust {L}{G}{M}{D} neuron}, Robotics and Autonomous Systems
  \textbf{30} (2000), 17--38.

\bibitem{CarpenterGrossberg81}
G.A. Carpenter and S.~Grossberg, \emph{Adaptation and transmitter gating in
  vertebrate photoreceptors}, Journal of Theoretical Biology \textbf{1} (1981),
  1--42.

\bibitem{CizekFaigl19}
P.~Cizek and J.~Faigl, \emph{Self-supervised learning of the
  biologically-inspired obstacle avoidance of hexapod walking robot},
  Bioinspiration \& Biomimetics \textbf{14} (2019), no.~4, 046002.

\bibitem{CoxMeyersSinha2004}
D.~Cox, E.~Meyers, and P.~Sinha, \emph{Contextually evoked object-specific
  responses in human visual cortex}, Science \textbf{304} (2004), no.~5667,
  115--117.

\bibitem{DemircigilEtAl2017}
M.~Demircigil, J.~Heusel, M.~Löwe, S.~Upgang, and F.~Vermet, \emph{On a model
  of associative memory with huge storage capacity}, Journal of Statistical
  Physics \textbf{168} (2017), no.~2, 288--299.

\bibitem{DoBaPa22}
C.G. Donohue, Z.M. Bagheri, J.C. Partridge, and J.M. Hemmi, \emph{Fiddler crabs
  are unique in timing their escape responses based on speed-dependent visual
  cues}, Current Biololgy \textbf{32} (2022), no.~23, 5159–5164.e4.

\bibitem{EykholtEtAl2018}
Kevin Eykholt, Ivan Evtimov, Earlence Fernandes, Bo~Li, Amir Rahmati, Chaowei
  Xiao, Atul Prakash, Tadayoshi Kohno, and Dawn Song, \emph{Robust
  physical-world attacks on deep learning visual classification}, 2018 IEEE/CVF
  Conference on Computer Vision and Pattern Recognition, 2018, pp.~1625--1634.

\bibitem{FuHuPe20}
Q.~Fu, C.~Hu, J.~Peng, F.C. Rind, and S.~Yue, \emph{A robust collision
  perception visual neural network with specific selectivity to darker
  objects}, IEEE Transactions on Cybernetics \textbf{50} (2020), no.~12,
  5074--5088.

\bibitem{FuHuPe18}
Q.~Fu, C.~Hu, J.~Peng, and S.~Yue, \emph{Shaping the collision selectivity in a
  looming sensitive neuron model with parallel {ON} and {OFF} pathways and
  spike frequency adaptation}, Neural Networks \textbf{106} (2018), 127--143.

\bibitem{GabKraKocLau02}
F.~Gabbiani, H.G. Krapp, C.~Koch, and G.~Laurent, \emph{Multiplicative
  computation in a visual neuron sensitive to looming}, Nature \textbf{420}
  (2002), 320--324.

\bibitem{GallegoEtAl2022}
G.~Gallego, T.~Delbrück, G.~Orchard, C.~Bartolozzi, B.~Taba, A.~Censi,
  S.~Leutenegger, A.J. Davison, J.~Conradt, K.~Daniilidis, and D.~Scaramuzza,
  \emph{Event-based vision: A survey}, IEEE Transactions on Pattern Analysis
  and Machine Intelligence \textbf{44} (2022), no.~1, 154--180.

\bibitem{Gibson1950}
J.J. Gibson, \emph{{T}he {P}erception of the {V}isual {W}orld}, Houghton
  Mifflin, Boston, 1950.

\bibitem{GotShi17}
Yuki Goto and Katsunari Shibata, \emph{Influence of the chaotic property on
  reinforcement learning using a chaotic neural network}, Neural Information
  Processing (Cham) (Derong Liu, Shengli Xie, Yuanqing Li, Dongbin Zhao, and
  El-Sayed~M. El-Alfy, eds.), Springer International Publishing, 2017,
  pp.~759--767.

\bibitem{Hartbauer17}
F.J. Harris, \emph{Simplified bionic solutions: A simple bio-inspired vehicle
  collision detection system}, Bioinspiration \& Biomimetics \textbf{12}
  (2017), no.~2, 026007.

\bibitem{HasRei56}
B.~Hassenstein and W.~Reichardt, \emph{Systemtheoretische {A}nalyse der
  {Z}eit-, {R}eihenfolgen- und {V}orzeichenauswertung bei der
  {B}ewegungsperzeption des {R}üsselk{\"a}fers {C}hlorophanus}, Zeitschrift
  für Naturforschung B \textbf{11} (1956), no.~9-10, 513--524.

\bibitem{HatsopoulosEtAl95}
N.~Hatsopoulos, F.~Gabbiani, and G.~Laurent, \emph{Elementary computation of
  object approach by a wide-field visual neuron}, Science \textbf{270} (1995),
  1000--1003.

\bibitem{HeAoWh20}
Lei He, Nabil Aouf, James~F. Whidborne, and Bifeng Song, \emph{Integrated
  moment-based {LGMD} and deep reinforcement learning for {UAV} obstacle
  avoidance}, 2020 IEEE International Conference on Robotics and Automation
  (ICRA), 2020, pp.~7491--7497.

\bibitem{Hopfield82}
J.J. Hopfield, \emph{Neural networks and physical systems with emergent
  collective computational abilities}, Proceedings of the National Academy of
  Sciences USA \textbf{79} (1982), no.~8, 2554–2558.

\bibitem{Hopfield84}
\bysame, \emph{Neurons with graded response have collective computational
  properties like those of two-state neurons}, Proceedings of the National
  Academy of Sciences USA \textbf{81} (1984), no.~10, 3088--3092.

\bibitem{Hu1962}
Ming-Kuei Hu, \emph{Visual pattern recognition by moment invariants}, IRE
  Transactions on Information Theory \textbf{8} (1962), no.~2, 179--187.

\bibitem{JudgeRind97}
S.~Judge and F.C. Rind, \emph{The locust dcmd, a movement-detecting neurone
  tightly tuned to collision trajectories}, Journal of Experimental Biology
  \textbf{200} (1997), no.~16, 2209--2216.

\bibitem{NIPS2011_0348}
M.S. Keil, \emph{Emergence of multiplication in a biophysical model of a
  wide-field visual neuron for computing object approaches: Dynamics, peaks, \&
  fits}, Advances in Neural Information Processing Systems 24 (J.~Shawe-Taylor,
  R.S. Zemel, P.~Bartlett, F.C.N. Pereira, and K.Q. Weinberger, eds.), 2011,
  pp.~469--477.

\bibitem{Mats2015}
\bysame, \emph{Dendritic pooling of noisy threshold processes can explain many
  properties of a collision-sensitive visual neuron}, PLoS Computational
  Biology \textbf{11} (2015), no.~10, e1004479.

\bibitem{MatsJoan2012}
M.S. Keil and J.~L{\'o}pez-Moliner, \emph{Unifying time to contact estimation
  and collision avoidance across species}, PLoS Computational Biology
  \textbf{8} (2012), no.~8, e1002625.

\bibitem{MatsEliAngel04}
M.S. Keil, E.~Roca-Morena, and A.~Rodr\'{\i}guez-V\'azquez, \emph{A neural
  model of the locust visual system for detection of object approaches with
  real-world scenes}, Proceedings of the Fourth IASTED International Conference
  (Marbella, Spain), vol. 5119, 6-8 September 2004, pp.~340--345.

\bibitem{MatsAngel03gc}
M.S. Keil and A.~Rodr\'{\i}guez-V\'azquez, \emph{Towards a computational
  approach for collision avoidance with real-world scenes}, Proceedings of
  SPIE: Bioengineered and Bioinspired Systems (Maspalomas, Gran Canaria, Canary
  Islands, Spain) (A.~Rodr\'{\i}guez-V\'azquez, D.~Abbot, and R.~Carmona,
  eds.), vol. 5119, SPIE - The International Society for Optical Engineering,
  19-21 May 2003, pp.~285--296.

\bibitem{KlapoetkeEtAl17}
N.C. Klapoetke, A.~Nern, M.Y. Peek, E.M. Rogers, P.~Breads, G.M. Rubin, M.B.
  Reiser, and G.M. Card, \emph{Ultra-selective looming detection from radial
  motion opponency}, Nature \textbf{551} (2017), 237–241.

\bibitem{KrotovHopfield2016}
D.~Krotov and J.J. Hopfield, \emph{Dense associative memory for pattern
  recognition}, Advances in Neural Information Processing Systems \textbf{29}
  (2016), 1172--1180.

\bibitem{KrotovHopfield2020}
\bysame, \emph{Large associative memory problem in neurobiology and machine
  learning}, arxiv.org \textbf{2008.06996 [q-bio.NC]} (2020).

\bibitem{LePeLi23}
F.~Lei, Z.~Peng, M.~Liu, J.~Peng, V.~Cutsuridis, and S.~Yue, \emph{A robust
  visual system for looming cue detection against translating motion}, IEEE
  Transactions on Neural Networks and Learning Systems \textbf{34} (2023),
  no.~11, 8362--8376.

\bibitem{GustavoElAl05}
G.~Li{\~n}{\'a}m, J.~Cuadri, M.S. Keil, R.~Stafford, and E.~Roca, \emph{A
  bio-inspired collision detection algorithm for {V}{L}{S}{I} implementation},
  Proceedings of SPIE: Bioengineered and Bioinspired Systems {I}{I} (Sevilla,
  Spain) (A.~Rodr\'{\i}guez-V\'azquez, E.~Roca, and D.~Abbot, eds.), SPIE - The
  International Society for Optical Engineering, 9-11 May 2005.

\bibitem{Mitchell96}
Melanie Mitchell, \emph{An introduction to genetic algorithms}, MIT Press,
  Cambridge, MA, 1996.

\bibitem{Palka67}
J.~Palka, \emph{An inhibitory process influencing visual responses in a fibre
  of the ventral nerve cord of locusts}, Journal of Insect Physiology
  \textbf{13} (1967), no.~2, 235–248.

\bibitem{RamsauerEtAl2020}
H.~Ramsauer, B.~Schäfl, J.~Lehner, P.~Seidl, M.~Widrich, L.~Gruber,
  M.~Holzleitner, M.~Pavlović, G.~Kjetil~Sandve, V.~Greiff, D.~Kreil, M.~Kopp,
  G.~Klambauer, J.~Brandstetter, and S.~Hochreiter, \emph{Hopfield networks is
  all you need}, arxiv.org \textbf{2008.02217 [cs.NE]} (2020).

\bibitem{Rind84}
F.C. Rind, \emph{A chemical synapse between two motion detecting neurones in
  the locust brain}, Journal of Experimental Biology \textbf{110} (1984),
  143--167.

\bibitem{RindBramwell96}
F.C. Rind and D.I. Bramwell, \emph{Neural network based on the input
  organization of an identified neuron signaling implending collision}, Journal
  of Neurophysiology \textbf{75} (1996), no.~3, 967--985.

\bibitem{RindSanter04}
F.C. Rind and D.R. Santer, \emph{Collision avoidance and a looming sensitive
  neuron: size matters but biggest is not necessarily best}, Proceeding of the
  Royal Society of London B \textbf{271} (2004), S27--S29.

\bibitem{RindSimmons92a}
F.C. Rind and P.J. Simmons, \emph{Orthopteran {DCMD} neuron: a reevaluation of
  responses to moving objects. {I}. {S}elective responses to approaching
  objects}, Journal of Neurophysiology \textbf{68} (1992), no.~5, 1654--1666.

\bibitem{RindSimmons98}
\bysame, \emph{Local circuit for the computation of object approach by an
  identified visual neuron in the locust}, The Journal of Comparative Neurology
  \textbf{395} (1998), 405--415.

\bibitem{RobJoh93b}
R.~M. Robertson and A.~G. Johnson, \emph{Retinal image size triggers obstacle
  avoidance in flying locusts}, Naturwissenschaften \textbf{80} (1993),
  176–178.

\bibitem{RowellEtAl77}
C.H.F. Rowell, M.~O'Shea, and J.L.D. Williams, \emph{The neuronal basis of a
  sensory analyser, the acridid movement detector system.{IV}.{T}he preference
  for small field stimuli}, Journal of Experimental Biology \textbf{68} (1977),
  157--185.

\bibitem{YueRind06}
Yue. S. and Rind. F.C., \emph{Collision detection in complex dynamic scenes
  using an {LGMD}-based visual neural network with feature enhancement}, IEEE
  Transactions on Neural Networks \textbf{17} (2006), no.~3, 705--716.

\bibitem{Scharr00}
H.~Scharr, \emph{Optimal operators in digital image processing ({O}ptimale
  {O}peratoren in der {D}igitalen {B}ildverarbeitung)}, Ph.D. thesis,
  University of Heidelberg (Germany), Combined Faculties for the Natural
  Sciences and for Mathematics, 2000.

\bibitem{Schlotterer1977}
G.R. Schlotterer, \emph{Response of the locust descending movement detector
  neuron to rapidly approaching and withdrawing visual stimuli}, Canadian
  Journal of Zoology \textbf{55} (1977), 1372--1376.

\bibitem{RindSimmons92b}
P.J. Simmons and F.C. Rind, \emph{Orthopteran {DCMD} neuron: a reevaluation of
  responses to moving objects. {II}. {C}ritical cues for detecting approaching
  objects}, Journal of Neurophysiology \textbf{68} (1992), no.~5, 1667--1682.

\bibitem{SimmonsRind1997}
\bysame, \emph{Responses to object approach by a wide field visual neurone the
  lgmd2 of the locust: characterization and image cues}, Journal of Comparative
  Physiology \textbf{180} (1997), no.~3, 203--214.

\bibitem{SitawarinEtAl18}
Chawin Sitawarin, Arjun Nitin~Bhagoji, Arsalan Mosenia, Mung Chiang, and
  Prateek Mittal, \emph{Darts: Deceiving autonomous cars with toxic signs},
  arXiv.org \textbf{1802.06430 [cs.CR]} (2018).

\bibitem{StGaMc19}
Emma Strubell, Ananya Ganesh, and Andrew McCallum, \emph{Energy and policy
  considerations for deep learning in {NLP}}, Proceedings of the 57th Annual
  Meeting of the Association for Computational Linguistics (Florence, Italy),
  Association for Computational Linguistics, July 2019, pp.~3645--3650.

\bibitem{OnePixelAttack2019}
Jiawei Su, Danilo~Vasconcellos Vargas, and Kouichi Sakurai, \emph{One pixel
  attack for fooling deep neural networks}, IEEE Transactions on Evolutionary
  Computation \textbf{23} (2019), no.~5, 828--841.

\bibitem{TayaraniSchmuker21}
M.H. Tayarani-Najaran and M.~Schmuker, \emph{Event-based sensing and signal
  processing in the visual, auditory, and olfactory domain: A review},
  Frontiers in Neural Circuits \textbf{31} (2021), no.~15, 610446.

\bibitem{TeufelDakinFletcher18}
C.~Teufel, S.C. Dakin, and P.C. Fletcher, \emph{Prior object-knowledge sharpens
  properties of early visual feature-detectors}, Scientific Reports \textbf{8}
  (2018), no.~1, 10853.

\bibitem{TomRin23}
D.~Tomsic and F.C. Rind, \emph{Animal behavior: Timing escape on angular size
  or angular velocity?}, Current Biology \textbf{33} (2023), no.~3, R108--R110.

\bibitem{VaShPa2017}
A.~Vaswani, N.~Shazeer, N.~Parmar, J.~Uszkoreit, L.~Jones, A.~N. Gomez, Ł.
  Kaiser, , and I.~Polosukhin, \emph{Attention is all you need}, Advances in
  Neural Information Processing Systems \textbf{30} (2017), 5998–6008.

\bibitem{WeScLe21}
Patrick Wenzel, Torsten Schön, Laura Leal-Taixé, and Daniel Cremers,
  \emph{Vision-based mobile robotics obstacle avoidance with deep reinforcement
  learning}, 2021 IEEE International Conference on Robotics and Automation
  (ICRA), 2021, pp.~14360--14366.

\bibitem{WidrichEtAl2020}
M.~Widrich, B.~Schäfl, M.~Pavlović, H.~Ramsauer, L.~Gruber, M.~Holzleitner,
  J.~Brandstetter, G.~Kjetil~Sandve, V.~Greiff, S.~Hochreiter, and
  G.~Klambauer, \emph{Modern hopfield networks and attention for immune
  repertoire classification}, Advances in Neural Information Processing Systems
  \textbf{33} (2020), 18832--18845.

\bibitem{YueEtAl06}
S.~Yue, F.C. Rind, M.S. Keil, J.~Cuadri-Carvajo, and R.~Stafford, \emph{A
  bio-inspired visual collision detection mechanism for cars: Optimisation of a
  model of a locust neuron to a novel environment}, Neurocomputing \textbf{69}
  (2006), no.~13–15, 1591--1598.

\bibitem{ZhLiKi22}
Baohua Zhou, Zifan Li, Sunnie Kim, John Lafferty, and Damon~A Clark,
  \emph{Shallow neural networks trained to detect collisions recover features
  of visual loom-selective neurons}, eLife \textbf{11} (2022), e72067.

\end{thebibliography}
%\bibliographystyle{amsplain}
%
% CONTENT FROM draft.bbl follows
%
\providecommand{\bysame}{\leavevmode\hbox to3em{\hrulefill}\thinspace}
\providecommand{\MR}{\relax\ifhmode\unskip\space\fi MR }
\providecommand{\MRhref}[2]{%
  \href{http://www.ams.org/mathscinet-getitem?mr=#1}{#2}
}
\providecommand{\href}[2]{#2}

}
\end{document}